\documentclass{article}

\usepackage[T1]{fontenc}
\usepackage[utf8]{inputenc}
\usepackage{textcomp}

\usepackage[english]{babel}
\usepackage{xspace}
\usepackage{setspace}
\usepackage[final]{microtype}
\usepackage{balance}

\usepackage{graphicx, tikz, epstopdf, capt-of, subfigure, stfloats, bbding, capt-of}
\usepackage{float}
\graphicspath{{figs/}}
\usepackage[colorinlistoftodos]{todonotes}
\usepackage{wrapfig}

\usepackage{amsmath, amsfonts, amssymb, nicefrac, pifont, siunitx}
\newcommand{\cmark}{\text{\ding{51}}}
\newcommand{\xmark}{\text{\ding{55}}}
\usepackage{algorithmic}

\DeclareMathAlphabet{\mathbbold}{U}{bbold}{m}{n}
\usepackage{booktabs, array, tabu, multirow, multicol}
\usepackage[flushleft]{threeparttable}

\usepackage[inline]{enumitem}
\usepackage{listings}
\usepackage{colortbl}

\usepackage[hyphens]{url}
\usepackage[labelfont=bf]{caption}
\usepackage{hyperref}
\usepackage{natbib}

\usepackage[]{cleveref} % get fancy referencing
\crefformat{section}{\S#2#1#3}
\crefrangeformat{section}{\S#3(#1)#4--\S#5(#2)#6}
\crefmultiformat{section}{\S#2#1#3}{ and~\S#2#1#3}{, \S#2#1#3}{, and~\S#2#1#3}
\crefformat{subsection}{\S#2#1#3}
\crefmultiformat{subsection}{\S#2#1#3}{, and~\S#2#1#3}{, \S#2#1#3}{, and~\S#2#1#3}
\crefformat{figure}{Fig.~#2#1#3}
\crefmultiformat{figure}{Figs.~#2#1#3}{, and~#2#1#3}{, #2#1#3}{, and~#2#1#3}
\crefformat{table}{Table~#2#1#3}
\crefrangeformat{table}{Tables~#3#1#4--#5#2#6}
\crefmultiformat{table}{Tables~#2#1#3}{, and~#2#1#3}{, #2#1#3}{, and~#2#1#3}
\crefformat{equation}{Eqn. #2#1#3}
\crefmultiformat{equation}{Eqns.~#2#1#3}{, and~#2#1#3}{, #2#1#3}{, and~#2#1#3}

\usepackage[accepted]{icml2020}

\icmltitlerunning{Inductive-bias-driven Reinforcement Learning for Efficient Schedules in Heterogeneous Clusters}

\begin{document}
    \twocolumn[
        \icmltitle{Inductive-bias-driven Reinforcement Learning for Efficient Schedules in Heterogeneous Clusters}
        \begin{icmlauthorlist}
        \icmlauthor{Subho S. Banerjee}{uiuc}
        \icmlauthor{Saurabh Jha}{uiuc}
        \icmlauthor{Zbigniew T. Kalbarczyk}{uiuc}
        \icmlauthor{Ravishankar K. Iyer}{uiuc}
        \end{icmlauthorlist}
        \icmlaffiliation{uiuc}{University of Illinois at Urbana-Champaign, USA}
        \icmlcorrespondingauthor{Subho S. Banerjee}{ssbaner2@illinois.edu}
        \icmlkeywords{Scheduling, Accelerators, POMDP, Bayesian, Sampling}
        \vskip 0.3in
    ]
    \printAffiliationsAndNotice{}

    % Macros
    \newcommand\Mark[1]{\textsuperscript{#1}}
    \newcommand*\circled[1]{#1}

    \begin{abstract}
    
The problem of scheduling of workloads onto heterogeneous processors (e.g., CPUs, GPUs, FPGAs) is of fundamental importance in modern data centers.
Current system schedulers rely on application/system-specific heuristics that have to be built on a case-by-case basis.
Recent work has demonstrated ML techniques for automating the heuristic search by using black-box approaches which require significant training data and time, which make them challenging to use in practice.
This paper presents Symphony, a scheduling framework that addresses the challenge in two ways:
\begin{enumerate*}[label=(\roman*)]
    \item a domain-driven Bayesian reinforcement learning (RL) model for scheduling, which inherently models the resource dependencies identified from the  system architecture; and
    \item a sampling-based technique to compute the gradients of a Bayesian model without performing full probabilistic inference.
\end{enumerate*}
Together, these techniques reduce both the amount of training data and the time required to produce scheduling policies that significantly outperform black-box approaches by up to 2.2\texttimes{}.

\end{abstract}
    \section{Introduction} \label{sec:intro}

The problem of scheduling of workloads on heterogeneous processing fabrics (i.e., accelerated datacenters including GPUs, FPGAs, and ASICs, e.g.,~\citet{Asanovic2014, Shao2015}), is at its core an intractable NP-hard problem~\cite{Mastrolilli2008, Mastrolilli2009}.
System schedulers generally rely on application- and system-specific heuristics with extensive domain-expert-driven tuning of scheduling policies (e.g.,~\citet{Isard2009,  Giceva2014, Lyerly2018, Mars2011, Mars2013, Ousterhout2013, Xu2018, Yang2013, Zhang2014, Zhuravlev2010, Zaharia2010}).
Such heuristics are difficult to generate, as variations across applications and system configurations mean that significant amounts of time and money must be spent in painstaking heuristic searches. 
Recent work has demonstrated machine learning (ML) techniques~\cite{Delimitrou2013, Delimitrou2014, Mao2016, Mao2018} for automating heuristic searches by using black-box approaches which require significant training data and time, making them challenging to use in practice.

This paper presents Symphony, a scheduling framework that addresses the challenge in two ways:
\begin{enumerate*}[label=(\roman*)]
    \item we use a domain-guided Bayesian-model-based partially observable Markov decision process (POMDP)~\cite{Astrom1965, Kaelbling1998} to decrease the amount of training data (i.e., sampled trajectories); and
    \item a sampling-based technique that allows one to compute the gradients of a Bayesian model without performing full probabilistic inference. 
\end{enumerate*}
We thus, significantly reduce the costs of (i) running a large heterogeneous computing system that uses an efficient scheduling policy; and (ii) training the policy itself.

\textbf{Reducing Training Data.}
State-of-the-art methods for choosing an optimal action in POMDPs rely on training of neural networks (NNs)~\cite{Mnih2016, Dhariwal2017}.
As these approaches are model-free, training of the NN requires large quantities of data and time to compute meaningful policies.
In contrast, we provide an inductive bias for the reinforcement learning (RL) agent by encoding domain knowledge as a Bayesian model that can infer the latent state from observations, while at the same time leveraging the scalability of deep learning methods through end-to-end gradient descent.
In the case of scheduling, our inductive bias is a set of statistical relationships between measurements from microarchitectural monitors~\cite{Dreyer1997}. 
To the best of our knowledge, this is the first paper to exploit those relationships and measurements to infer resource utilization in the system (i.e., latent state) to build RL-based scheduling polices.

\textbf{Reducing Training Time.}
The addition of the inductive bias, while making the training process less data-hungry (i.e., requiring fewer workload executions to train the model), comes at the cost of additional training time: the cost of performing full-Bayesian inference at every training step~\cite{Dagum1993, Russell1995, Binder1997}.
It is this cost that makes the use of deep RL techniques in dynamic real-world deployments (which require periodic retraining) prohibitively expensive.
To address that issue, we have developed a procedure for computing the gradient of variables in the above Bayesian model without requiring full inference computation, unlike prior work~\cite{Russell1995, Binder1997}. 
The key is to calculate the gradient by generating samples from the model, which is computationally simpler than inferring the posterior distribution.

\begin{figure}[!t]
    \centering
    \includegraphics[width=\columnwidth]{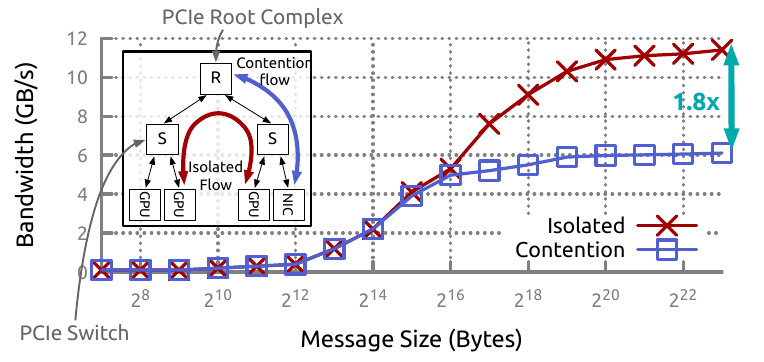}
    \vspace{-.8cm}
    \caption{Performance degradation due to PCIe contention between GPU and NIC (averaged over 10 runs).}
    \label{fig:pcie_perf}
\end{figure}

\textbf{Need for New Scheduler.}
Current schedulers prioritize the use of simple generalized heuristics and coarse-grained resource bucketing (e.g., core counts, free memory) to make scheduling decisions. 
Hence, even though they are perceived to perform well in practice, they do not model complex emergent heterogeneous compute platforms and hence leave a lot to be desired.
Consider the case of a distributed data processing framework that uses two GPUs to perform a \textit{halo exchange}.\footnote{A \textit{halo exchange} occurs due to communication arsing between parallel processors computing an overlapping pieces of data, called \textit{halo regions}, that need to be periodically updated.}
\cref{fig:pcie_perf} shows the performance (here, bandwidth) of the exchange as ``isolated'' performance.
If the application were to concurrently perform distributed network communication, we would observe that the original GPU-to-GPU communication is affected because of PCIe bandwidth contention at shared links (i.e., a ``hidden'' resource that is not often exposed to the user).
Such behavior is shown as ``contention'' in \cref{fig:pcie_perf}, and can cause as much as a $0-1.8\times$ slowdown, depending on the size of the transmitted messages.
Traditional approaches would either have such a heuristic manually searched and incorporated into a scheduling policy, or would expect it to be found automatically as part of the training of a black-box ML model, and both approaches can require significant effort in profiling/training.
In contrast, our approach allows the utilization of architectural resources (in this case, of the PCIe network) as an inductive bias for the RL-agent, thereby allowing the training process to automatically hone in on such resources of interest, without having to identify the resource's importance manually.

\textbf{Results.}
The Symphony framework reduces the average job completion time over hand-tuned scheduling heuristics by as much as 32\%, and to within 6\% of the time taken by an oracle scheduler.
It also achieves a training time improvement of 4\texttimes{} compared to full Bayesian inference based on belief propagation.
Further, the technique outperforms black-box ML techniques by 2.2\texttimes{} in terms of training time. 
We believe that Symphony is also representative of RL applied to several other control-related problems (e.g., industrial scheduling, data center network scheduling) where data-driven approaches can be augmented with domain knowledge to build sample-efficient RL-agents.
    \section{Background} \label{sec:background}

\textbf{Partially Observable Markov Decision Processes.}
A POMDP is a stochastic model that describe relationships between an agent and its environment. It is a tuple $(\mathcal{S}, \mathcal{A}, \mathcal{T}, \Omega, O, R, \gamma)$, where $\mathcal{S}$ is the state space, $\mathcal{A}$ is the action space, and $\Omega$ is the observation space.
We use $s_t \in \mathcal{S}$ to denote the hidden state at time $t$.
When an action $a_t \in \mathcal{A}$ is executed, the state changes according to the transition distribution, $s_{t+1} \sim \mathcal{T}(s_{t+1} | s_t, a_t)$.
Subsequently, the agent receives a noisy or partially occluded observation $o_{t+1} \in \Omega$ according to the distribution
$o_{t+1} \sim O(o_{t+1} | s_{t+1}, a_t)$, and a reward $r_{t+1} \in \mathbb{R}$ according to the distribution $r_{t+1} \sim R(r_{t+1} | s_{t+1}, a_t)$.

An agent acts according to its policy $\pi(a_t | s_t)$, which returns the probability of taking action $a_t$ at time $t$.
The agent's goal is to learn a policy $\pi$ that maximizes the expected future reward $J = \mathbb{E}_{\tau \sim p(\tau)}[\sum_{t=1}^{T} \gamma^{t-1}r_t]$ over trajectories $\tau = (s_0, a_0, \dots, a_{T-1}, s_T)$ induced by its policy, where $\gamma \in [0, 1)$ is the discount factor.
In general, a POMDP agent must infer the belief state $b_t = \Pr(s_t |  o_1, \dots, o_t, a_0, \dots, a_{t-1})$, which is used to calculate $\pi(a_t | \hat s_t)$ where $\hat s_t \sim b_t$. In the remainder of the paper, we will use $\pi(a_t | \hat s_t)$ and $\pi(a_t | b_t)$ interchangeably.

\textbf{Related Work.}
Finding solutions for many POMDPs involves
\begin{enumerate*}[label=(\roman*)]
    \item estimating the transition model $T$ and observation model $O$,
    \item performing inference under this model, and 
    \item choosing an action based on the inferred belief state.
\end{enumerate*}
Prior work in this area has extensively explored the use of NNs, particularly recurrent NNs (RNNs), as universal function approximators for (i) and (iii) above because they can be easily trained and have efficient inference procedures (e.g.,~\citet{Hausknecht2015, Narasimhan2015, Mnih2015, Jaderberg2016, Foerster2016, Karkus2017, Zhu2018}).
Neural networks have proven to be extremely effective at learning, but usually require a lot of data (for RL-agents, sampled trajectories, which may be prohibitively expensive to acquire for certain classes of applications, such as scheduling).
The ability to incorporate explicit domain knowledge (which in the case of scheduling, is based on system design invariants) could significantly reduce the amount of data required.
To that end, other work~\cite{Karkus2017,Silver2017,Igl2018} has advocated the integration of probabilistic models (including Bayesian filter models) for (i) above.
The significant computational cost of learning and inference in such deep probabilistic models has spurred the use of approximation techniques for training and inference, including NN-based approximations of Bayesian inference~\cite{Karkus2017, Zhu2018} and variational inference methods~\cite{Igl2018}.

In this paper, we too advocate the use of a domain-driven probabilistic model for $b_t$ that can be trained through end-to-end back-propagation to compute a policy.
Specifically, the technique handles the gradient descent procedure on a Bayesian network (BN) with known structure and incomplete observations without performing inference on the BN, only requiring generation of samples from the model.
That approach is different from to prior work on learning BNs using gradient descent~\cite{Russell1995, Binder1997} or expectation maximization, both of which require full posterior inference at every training step.

\textbf{Actor-Critic Methods.}
Actor-Critic methods~\cite{Konda2000} have previously been proposed for learning the parameters $\rho$ of an agent's policy $\pi_\rho(a_t|s_t)$.
Here 
\begin{enumerate*}[label=(\roman*)]
    \item the ``Critic'' estimates the value function $V(s)$, and 
    \item the ``Actor'' updates the policy $\pi(a | s)$ in the direction suggested by the Critic.
\end{enumerate*}
In this paper, we use $n$-step learning with the asynchronous advantage actor-critic (A3C) method~\cite{Mnih2016}.
For $n$-step learning, starting at time $t$, the current policy performs $n_s$ consecutive steps in $n_e$ parallel environments.
The gradient updates of $\pi$ and $V$ are based on that mini-batch of size $n_en_s$.
The target for the value function $V_\eta(s_{t+i})$, $i \in [0, n_s)$, parameterized by $\eta$, is the discounted sum of on-policy rewards up until $t + n_s$ and the off-policy bootstrapped value $V^*_\eta(s_{t + n_s})$.
If we use an advantage function $A^{t, i}_\eta = 
(\sum_{j = 0}^{n_s - i - 1} \gamma^j r_{t + i + j}) + \gamma^{n_s - i}V^*_\eta(s_{t + n_s}) - V_\eta(s_{t+1})$, the value function is
\begingroup
\allowdisplaybreaks
\begin{subequations}
    \begin{align}
        \mathcal{L}^A_t(\rho) &= -\frac{1}{n_en_s} \sum_{e=0}^{n_e-1} \sum_{i = 0}^{n_s-1} \mathbb{E}_{s_{t+i}\sim b_{t+i}}[\log \pi_\rho(a_{t+i} | s_{t+i}) \nonumber \\&~ \hspace{3cm} A_\eta^{t, i}(s_{t+i}, a_{t+i}) ]\\
        \mathcal{L}^V_t(\eta) &= \frac{1}{n_en_s} \sum_{e=0}^{n_e-1} 
        \sum_{i = 0}^{n_s-1} \mathbb{E}_{s_{t+i}\sim b_{t+i}}\left[ A_\eta^{t, i}(s_{t+i}, a_{t+i})^2\right].
    \end{align}
    \label{eqn:loss_funcs}
\end{subequations}
\endgroup

    \vspace{-1cm}
\section{Training the POMDP RL-Agent with Back-Propagation} \label{sec:model}
\begin{figure}[!t]
    \centering
    \includegraphics[width=0.9\columnwidth]{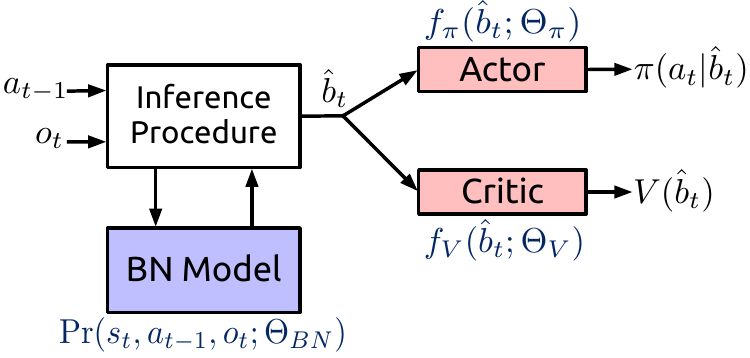}
    \vspace{-0.3cm}
    \caption{The proposed RL architecture.}
    \label{fig:model_grad}
\end{figure}

We consider a special case of the POMDP formulation presented above (illustrated in \cref{fig:model_grad}).
We assume that the domain knowledge about the environment of the RL-agent is presented as a joint probability distribution $\Pr(s_t, a_{t-1}, o_t; \Theta_{BN})$ that can be factorized as a BN (with parameters $\Theta_{BN}$). 
A BN is a probabilistic graphical model that represents a set of variables and their conditional dependencies via a directed acyclic graph (DAG).
We use probabilistic inference on the BN to calculate an estimate of the belief state $\hat b_t$.
$\hat b_t$ is then used in an NN $f_\pi(\hat b_t; \Theta_\pi)$ (with parameters $\Theta_\pi$) to  approximate the RL-agent's policy, and an NN $f_V(\hat b_t; \Theta_V)$ (with parameters $\Theta_V$) to  approximate the state-based value function.
We refer to all the parameters of the model as $\mathbf{\Theta} = (\Theta_{BN}, \Theta_\pi, \Theta_V) = (\rho, \eta)$.
The model is then trained by propagating the gradient of the total loss $\nabla_\mathbf{\Theta} \mathcal{L}_t^{RL} = \nabla_\mathbf{\Theta}\mathcal{L}_t^A(\rho) + \nabla_\mathbf{\Theta}\mathcal{L}_t^V(\eta)$. 
Estimating this gradient requires us to compute $\nabla_{\Theta_{BN}} \hat b_t$.
Traditional methods for computing the gradient require inference computation~\cite{Russell1995, Binder1997}.
However, even approximate inference in such models is known to be NP-Hard~\cite{Dagum1993}.
Below we describe an algorithm for approximating the gradient without requiring computation of full Bayesian inference.
All that is required is the ability to generate samples from the BN.
Only the subset of the BN necessary for generation of the samples is expanded.
The samples are then used as a representation of the distribution of the BN.
As a result, the proposed method decouples the training of the BN from the inference procedure used on it to calculate $\hat b_t$.

\subsection{The Bayesian Network \& Its Gradient}
Let the BN described above be a DAG $(V,E)$, and let $\mathbf{X} = \{X_v | v \in V \}$ be a set of random variables indexed by $V$.
Associated with each node $X$ is a conditional probability density function $\Pr(X | \wp(X))$, where $\wp(X)$ are the parents of $X$ in the graph. 
We assume that we are given
\begin{enumerate*}[label=(\roman*)]
    \item an efficient algorithm for sampling values of $X$ given $\wp(X)$, and
    \item a function $f_X(x, y; \theta_X) = \Pr_{\theta_X}(X = x | \wp(X) = y)$ whose partial derivative with respect to $\theta_X$ is known and efficiently computable. 
\end{enumerate*}
The BN can also have deterministic relationships between two random variables, under the assumption that the relationship is a differentiable diffeomorphism.
That is, for random variables $X$, $Y$, and diffeomorphism $F$, $\Pr(Y=y) = \Pr(X = F^{-1}(y))|DF^{-1}(y)|$ where $DF^{-1}$ is the inverse of the Jacobian of $F$.

\textbf{Computing Gradient.}
For a random variable $X$ in the BN, we define its parents as $\wp(X)$, its ancestor set as $\Xi(X) = \{Y | Y \rightsquigarrow X~\land~ Y \not\in \wp(X) \}$ (where $\rightsquigarrow$ represents a directed path in the BN). 
We now define a procedure to approximately compute the gradient of $X$ with respect to $\Theta_{BN}$.
We do so in two parts:
\begin{enumerate*}[label=(\roman*)]
    \item $\nicefrac{\partial \Pr(X = x | \xi = \mathbf{a})}{\partial \theta_X}$ and
    \item $\nabla_{\Theta_{BN}\setminus\theta_X} \Pr(X = x | \xi = \mathbf{a})$ for $\xi \subseteq \Xi(X)$.
\end{enumerate*}
First,
\begingroup
\allowdisplaybreaks
\begin{align}
    &\frac{\partial \Pr(X = x | \xi = \mathbf{a})}{\partial \theta_X}\nonumber \\
    &= \frac{\partial}{\partial \theta_X} \int \Pr(\wp(X) = \mathbf{y} | \xi = \mathbf{a}) \times \nonumber \\
    &~ \hspace{2cm} \Pr(X = x | \wp(X) = \mathbf{y}, \xi = \mathbf{a}) d\mathbf{y} \nonumber \\
    &= \frac{\partial}{\partial \theta_X} \int \Pr(\wp(X) = \mathbf{y} | \xi = \mathbf{a}) f_X(x, \mathbf{y}; \theta_X) d\mathbf{y}  \nonumber \\
    &= \int \Pr(\wp(X) = \mathbf{y} | \xi = \mathbf{a})  \frac{\partial f_X(x, \mathbf{y}; \theta_X)}{\partial \theta_X}  d\mathbf{y} \nonumber \\
    &\approx \sum_{i=1}^{S} \frac{\mathfrak{n}_S(\mathbf{a},\mathbf{y}_i)}{\mathfrak{n}_S(\mathbf{a})} \frac{\partial f_X(x, \mathbf{y}_i; \theta_X)}{\partial \theta_X}.
    \label{eqn:approx_grad_1}
\end{align}
\endgroup
Here, $S$ samples are drawn from a variable(s) $Z$ such that $\mathfrak{n}_S(j)$ is the number of times the value $j$ appears in the set of samples $\{z_i\}$, i.e., $\mathfrak{n}_S(j) = \sum_{i=1}^{S}\mathbbold{1}\{z_i = j\}$.
Next, 
\begin{align}
    &\nabla_{\Theta_{BN}\setminus\theta_X} \Pr(X = x | \xi = \mathbf{a}) \nonumber \\
    &= \nabla_{\Theta_{BN}\setminus\theta_X} \int \Pr(\wp(X) = \mathbf{y} | \xi = \mathbf{a}) \times \nonumber \\
    &~ \hspace{2cm}\Pr(X = x | \wp(X) = \mathbf{y}, \xi = \mathbf{a}) d\mathbf{y} \nonumber \\
    % &= \nabla_{\Theta_{BN}\setminus\theta_X} \int \Pr(\wp(X) = \mathbf{y} | \xi = \mathbf{a}) f_X(x, \mathbf{y}; \theta_X) d\mathbf{y} 
    &= \int f_X(x, \mathbf{y}; \theta_X) \nabla_{\Theta_{BN}\setminus\theta_X} \Pr(\wp(X) = \mathbf{y} | \xi = \mathbf{a})  d\mathbf{y} \nonumber \\
    &\approx \sum_{i=1}^{S} \frac{\mathfrak{n}_S(\mathbf{y}_i)}{S} f_X(x, \mathbf{y}_i; \theta_X) \nabla_{\Theta_{BN}\setminus\theta_X} \Pr(\wp(X) = \mathbf{y}_i | \xi = \mathbf{a}) 
    \label{eqn:approx_grad_2}
\end{align}
When $|\wp(X)| > 1$, variables in $\wp(X)$ might not be conditionally independent given $\Xi(X)$.
Hence we find a set of nodes $N$ such that $I \perp J | \Xi(X) \cup N ~\forall I, J \in \wp(X)$. 
Then,
\begin{align}
    &\Pr(\wp(X)=\mathbf{y}_i | \xi = \mathbf{a}) \nonumber \\
    &= \int \Pr(N=\mathbf{n} | \xi = \mathbf{a})Pr(\wp(X) = \mathbf{y} | N=\mathbf{n}, \xi = \mathbf{a}) d\mathbf{n} \nonumber \\
    &= \int \Pr(N=\mathbf{n} | \xi = \mathbf{a}) \prod_{j=1}^{m} \Pr(P_j = y_j | N=\mathbf{n}, \xi = \mathbf{a}) d\mathbf{n} \nonumber \\
    &\approx \sum_{k=1}^{S} \frac{\mathfrak{n}_S(\mathbf{a}, \mathbf{n}_i)}{\mathfrak{n}_S(\mathbf{a})} \prod_{j=1}^{m} \Pr(P_j = y_j | N=\mathbf{n}_k, \xi = \mathbf{a}),
    \label{eqn:approx_grad_3}
\end{align}
where $\wp(X) = (P_1, \dots, P_m)$ and $\mathbf{y}_i = (y_{i, 1}, \dots, y_{i, m})$. 
Thus, we obtain,
\begingroup
\allowdisplaybreaks
\begin{align}
    &\nabla_{\Theta_{BN}\setminus\theta_X} \Pr(\wp(X) = \mathbf{y}_i | \xi = \mathbf{a}) \nonumber \\
    &\approx \sum_{k=1}^{S} \frac{\mathfrak{n}_S(\mathbf{a}, \mathbf{n}_k)}{\mathfrak{n}_S(\mathbf{a})}\times \nonumber\\
    &~ \hspace{1cm} \nabla_{\Theta_{BN}\setminus\theta_X} \prod_{j=1}^{m} \Pr(P_j = y_{i,j} | N=\mathbf{n}_k, \xi = \mathbf{a}) \nonumber \\
    &= \sum_{k=1}^{S} \frac{\mathfrak{n}_S(\mathbf{a}, \mathbf{n}_k)}{\mathfrak{n}_S(\mathbf{a})}
    \times \nonumber \\ 
    &~ \hspace{1cm} \sum_{l=1}^{m}\left( \prod_{h=1, h\neq l}^m \Pr(P_h = y_{i, h} | N=\mathbf{n_k}, \xi = \mathbf{a})\right) \times \nonumber \\ 
    &~ \hspace{2.5cm} \nabla_{\Theta_{BN}\setminus\theta_X} \Pr(P_l = y_{i, l} | N=\mathbf{n}_k, \xi = \mathbf{a}) \nonumber \\
    &\approx \sum_{k=1}^{S} \frac{\mathfrak{n}_S(\mathbf{a}, \mathbf{n}_k)}{\mathfrak{n}_S(\mathbf{a})}
    \sum_{l=1}^{m} \left( \prod_{h=1, h\neq l}^m \frac{\mathfrak{n}_S(y_{i, h}, a, \mathbf{n}_k)}{\mathfrak{n}_S(a, \mathbf{n}_k)} \right) \times \nonumber \\
    &~ \hspace{1.3cm} \overbrace{\nabla_{\Theta_{BN}\setminus\theta_X} \Pr(P_l = y_{i,l} | N=\mathbf{n}_k, \xi = \mathbf{a})}^{\text{Expand by recursion using \cref{eqn:approx_grad_1,eqn:approx_grad_2,eqn:approx_grad_4}}}. 
    \label{eqn:approx_grad_4}
\end{align}
\endgroup
The term $\nabla_{\Theta_{BN}\setminus\theta_X} \Pr(P_l = y_{i,l} | N=\mathbf{n}_k, \xi = \mathbf{a})$ represents the gradient operator on a subset of the original BN, containing only the ancestors (from the BN's graphical structure) of $X$.
Hence that gradient term can be recursively expanded using \cref{eqn:approx_grad_1,eqn:approx_grad_2,eqn:approx_grad_4}.
Repeating that process for all variables in $\hat b_t$ allows us to calculate the $\nabla_{\Theta_{BN}} \hat b_t$.

\textbf{Computational Complexity.}
The cost of computing \cref{eqn:approx_grad_1,eqn:approx_grad_2} is $O(S)$.
The cost of computing \cref{eqn:approx_grad_4} is $O(mS)$.
The cost of finding $N$ is $O(|\wp(s_t)|^2(|V| + |E|))$ (i.e., the cost of running the Bayes ball algorithm~\cite{Shachter2013} for every pair of nodes in $\wp(X)$). 
The total computational complexity of the entire procedure hinges on finding the number of times \cref{eqn:approx_grad_1,eqn:approx_grad_3,eqn:approx_grad_4} are executed, which we refer to as $Q$. 
$Q$ depends on the size of $N$ and on the graphical structure of the BN.
Hence, the total cost of computing $\nabla_{\Theta_{BN}} \hat b_t$ is $O(Q(|\wp(s_t)|^2(|V| + |E|) + mS))$ (where $|\wp(s_t)|\leq |V|-1$), which is computed $n_sn_e|b_t|$ times during training.
Note that for a polytree BN (the graphical structure of the BN we will use in \cref{sec:scheduling}), $N = \varnothing$, and $Q \leq |V|$.
This is still better than belief propagation on the polytree with the gradient computation technique from~\citet{Russell1995, Binder1997}, which is $O(|V|\max_{v\in V}(dom(X_v)))$, where $dom(X)$ is the size of the domain of $X$, which could be exponentially large.

    \section{Scheduling Data Center Workloads By Using Reinforcement Learning} \label{sec:scheduling}
We now demonstrate an application of the POMDP model and training methodology presented in \cref{sec:model} to the problem of scheduling tasks on a heterogeneous processing fabric that includes CPUs, GPUs, and FPGAs.
The model integrates real-time performance measurements, prior knowledge about workloads, and system architecture to 
\begin{enumerate*}[label=(\roman*)]
    \item dynamically infer system state (i.e., resource utilization), and 
    \item automatically schedule tasks on a heterogeneous processing fabric.
\end{enumerate*}

\textbf{Workload \& Programming Model.}
The system workload consists of multiple user programs, and each program is expressed as a \emph{data flow graph} (DFG).
A DFG is a DAG where the nodes represent computations (which we refer to as \emph{kernels}, e.g., matrix multiplication), and edges represent input-output relationships between the nodes.
Prior work has shown that a large number of applications can be expressed as compositions of such kernels~\cite{Asanovic2009, Banerjee2016}.
Prominent examples of such compositions include modern data analytics and ML frameworks that describe workloads as DFGs~\cite{Abadi2016, Chambers2010, McCool2012, Zaharia2012}.
We assume that the kernels are known ahead of time and have multiple implementations available for different processors and accelerators.
That assumption is correct for many ML workloads; for other workloads, it is an area of active research wherein accelerator designers and architects are trying to decompose larger applications into smaller pieces.
Once trained, our approach can schedule any composition (DFG) of the kernels, but requires retraining when the set of available kernels change.

\begin{figure}[!t]
    \centering
    \includegraphics[width=\columnwidth]{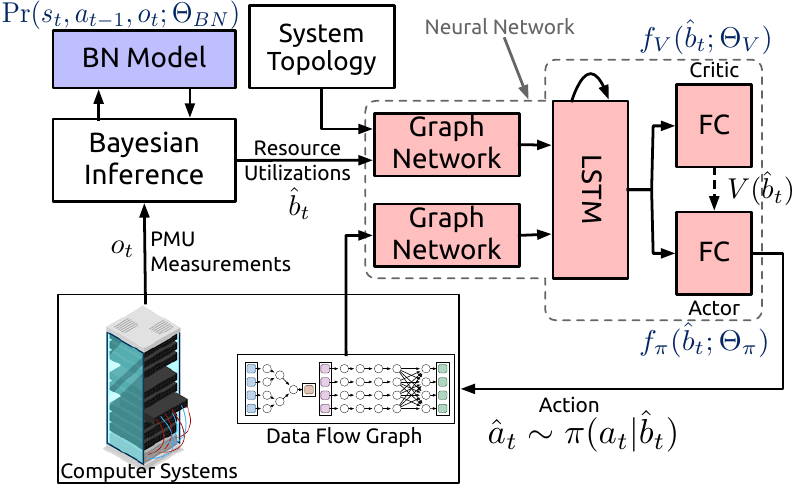}
    \vspace{-0.75cm}
    \caption{Architecture of the Symphony ML model.}
    \label{fig:scheduler}
\end{figure}

\textbf{POMDP Architecture.}
The overall architecture of the Symphony POMDP model is illustrated in \cref{fig:scheduler}.
The first part of the POMDP models the latent state $\hat b_t$ of the computer system. 
For the scheduling problem, $\hat b_t$ corresponds to resource utilization of various components of the computer system.
Utilization of some of the resources can be measured directly in software (e.g., the amount of free memory); however, the different layers of abstraction of the computer stack hide some others from direct measurement.
For example, consider the example in \cref{fig:pcie_perf} in \cref{sec:intro}; here, PCIe link bandwidth cannot be directly measured.
However, it can be measured indirectly by using the number of outstanding requests to memory from each PCIe device and by using the topology of the PCIe network.
In essence, we statistically relate the back pressure of one resource on another, until we can find a resource that can be directly measured via real-time performance counter (PC) measurements ($o_t$)~\cite{Dreyer1997}. 
We refer to such resources whose utilization cannot be directly measured as \textit{hidden} resources.
PCs are special-purpose registers present in the CPU and other accelerators for characterization of an application's behavior and identification of microarchitectural performance bottlenecks.
Specifically, we use a BN to 
\begin{enumerate*}[label=(\roman*)]
    \item model aleatoric uncertainty in measurements, and
    \item encode our knowledge about system architecture in terms of invariants or statistical relationships between the measurements.
\end{enumerate*} 
Inference on that BN then gives us an accurate estimate of the latent state of the system.
Second, we use an RNN (i.e., $f_\pi(\cdot)$ and $f_V(\cdot)$) to learn scheduling policies for user programs that minimize resource contention and maximize performance.
Those two ML models effectively decouple system-architecture-specific and measurement-specific aspects of scheduling (the BN) from its optimization aspects (the NN).
The compelling value of the above architecture (and its two constituent models) is that it can automatically generate scheduling policies for the deployment of DFGs in truly heterogeneous environments (that have CPUs, GPUs, and FPGAs) without requiring configuration specifics, or painstakingly tuned heuristics.
The model improves overall performance and resource utilization, and enables fine-grained resource sharing across workloads.

\begin{figure}[!t]
    \centering
    \includegraphics[width=\columnwidth]{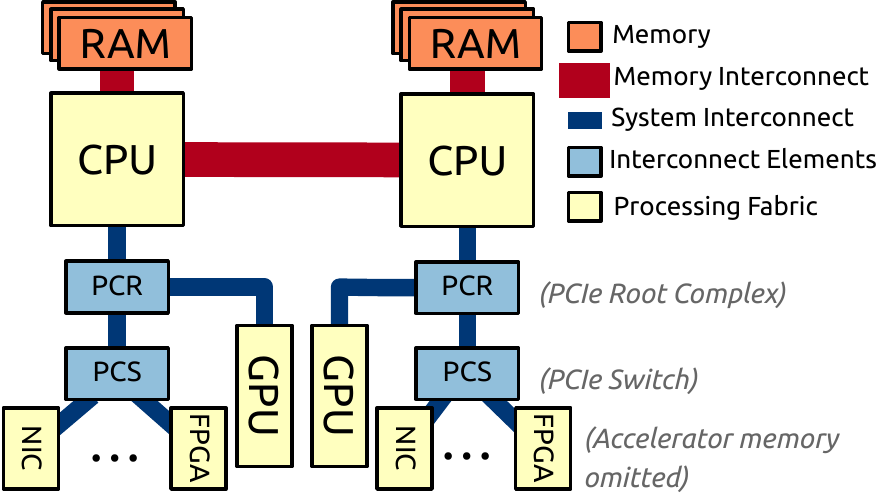}
    \vspace{-.7cm}
    \caption{Organization of a multi-CPU computer.}
    \label{fig:system_topology}
\end{figure}

\begin{figure}[!t]
    \centering
    \includegraphics[width=\columnwidth]{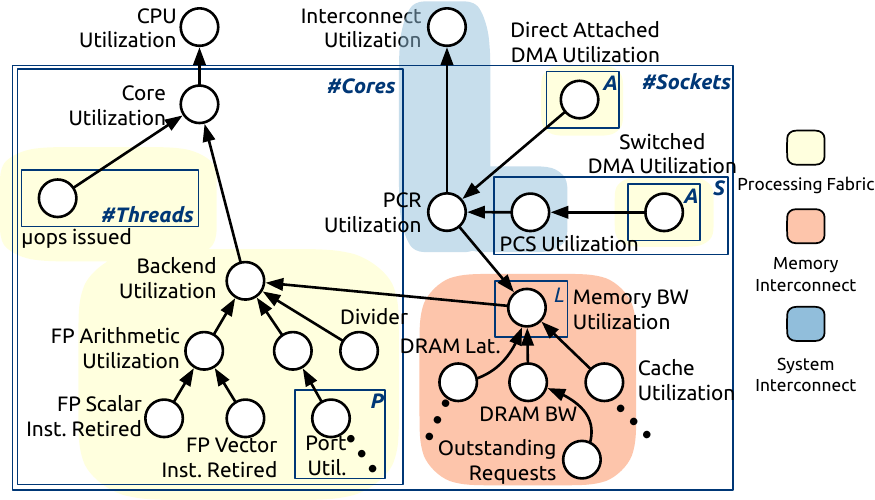}
    \vspace{-.7cm}
    \caption{Bayesian network (uses the plate notation) used to estimate resource utilization.}
    \label{fig:bn}
\end{figure}

\textbf{Performance Counters.} 
PCs are generally relied upon to conduct low-level performance analysis or tuning of performance bottlenecks in applications.
As the source of such bottlenecks is generally the unavailability of system resources, the performance counter can naturally be used to estimate resource utilization of a system.
Another benefit of using PCs is that it is not necessary to modify an application's source code in order to make measurements.
PCs can be grouped into three categories:
\begin{enumerate*}[label=(\roman*)]
    \item those pertaining to the processing fabric (CPU core or accelerators);
    \item those pertaining to the memory subsystem; and
    \item those pertaining to the system interconnect (in our case, PCIe).
\end{enumerate*}
\cref{fig:system_topology} illustrates the organization of a computer system as well as the categories above.
\cref{fig:bn} shows a mapping between the system organization and the PCs that are used in the BN model (described below).\footnote{A complete list of the PCs used in this paper can be found in the supplementary material.}

\textbf{BN Model.}
Measurements made from PCs have some inherent noise~\cite{Weaver2008}. 
The measurements can only be stored in a fixed number of registers.
Hence, only a fixed number of measurements can be made at any one point in time.
As a result, one must make successive measurements that capture marginally different system states.
Particular performance counters might become unavailable (or return incorrect values).
Finally, if a single scheduling agent is controlling a cluster of machines (which is common in data centers), measurements made on different machines will not be in sync and will often be delayed by network latency.
As a result, PCs are often sampled $N$ times between successive scheduler invocations to get around some of the sources of error.
To maximize the performance estimation fidelity, we apply statistical methods to systematically model the variance of the measurements.
For a single performance counter $o_t[c]$, if the error in measurement $e_c$ can be modeled, then the measured value $m_c$ can be modeled in terms of the true value $v_c$ plus measurement noise $e_c$, i.e., $m_c = v_c + e_c$.
Here, we focus only on random errors, and assume zero systematic error.
That is a valid assumption because the only reason for systematic errors is hardware or software bugs. 
We assume that the error can be modeled as $e_c \sim \mathcal{N}(0, \sigma)$ for some unknown variance $\sigma$; hence, $\Pr(m_c \mid v_c) = \mathcal{N}(m_c, \sigma)$.
That follows from prior work based on extensive measurement studies~\cite{Weaver2008}.
Now, given $N$ measurements of the value of the performance counter, we compute their sample mean $\mu$ and sample variance $S$.
A scaled and shifted t-distribution describes the marginal distribution of the unknown mean of a Gaussian, when the dependence on variance has been marginalized out~\cite{Gelman1995}; i.e., $v_c \sim \mu + \nicefrac{S}{\sqrt{N}}~Student(\nu = N-1).$
In our experiments, the confidence level of the t-distribution was 95\%.

Now, given a distribution of $v_c$ for every element of $o_t$, we describe the construction of the BN model.
Our goal is to model resource utilization (a number in $[0,1]$) for a relevant set of architectural resources $R$.
To do so, we use algebraic models for composing PC measurements ($v_c$) by using algebraic (deterministic) relationships derived from information about the CPU architecture.
Processor performance manuals~\cite{Yasin2014,IntelSDM,IBM2017_Perf} and or vendor contributions in OS codebases (e.g., in the \texttt{perf} module in Linux) provide such information.
When available in the later format (which is indeed the
case for all modern Intel, AMD, ARM, and IBM CPUs), these relationships can be automatically parsed and be used to construct the BN.

As our error-corrected measurements are defined in terms of distributions, the algebraic models that encode static information about relationships (based on the microarchitecture of the processor or topology of the system) now define statistical relationships $v_c$s (based on the Jacobian relationships described in \cref{sec:model}). 
\cref{fig:bn} shows an example of the BN model.
However, the types and meanings of hardware counters vary from one kind of architecture to another because of the variation in hardware organizations.
As a result, the model defined by the BN is parametric, changing with different processors and system topologies (i.e., across all the different types of systems in a data center).

Consider the example of identifying memory bandwidth utilization for a CPU core.
According to the processor documentation, the utilization can be computed by measuring the number of outstanding memory requests (which is available as a PC), i.e., $\nicefrac{\text{Outstanding Requests}[\geq \theta_{MB}]}{\text{Outstanding Requests}[\geq 1]}$.\footnote{Here $\text{X}[\geq t]$ counts cycles in which X exceeds threshold $t$.}
That is, identify the fraction of cycles in some time window that
CPU-core stalls because of insufficient bandwidth.
Naturally, in order to sustain maximum
performance, it is necessary to ensure that no stalls occur.
The value $\theta_{MB}$ is processor-specific and might not always be known.
In such cases, we use the training approach described in \cref{sec:model} to learn $\theta_{MB}$.
The procedure is repeated for all relevant system utilization counters (marked as ``Util.'' in \cref{fig:bn}), which together represent $\hat b_t$.
Such a BN model for a 16-core Intel Xeon processor (with all PCIe lanes populated) has 68 nodes, of which 32 are directly measured and the remainder are computed through inference.

\textbf{BN Retraining.}
The architectural information required to build the BN can be found in processor manuals~\cite{IntelSDM, IBMDMAPerf, IBM2017_Perf} as well as in machine-parsable databases in the Linux kernel source code as part of the \texttt{perf} package.
The only human intervention required in the process of building the BN is for filtering out those resources that cannot be controlled with software (because they change too quickly).
The BN model should only be rebuilt when the underlying hardware configuration changes, which \citet{Mars2013} observe happens every 5--6 years in a data center. 

\begin{table}[!t]
    \centering
    \caption{Performance counters used in test evaluation. We have disambiguated the names to ensure platform independence.}
    \label{tab:pmu_counters}
    \resizebox{\columnwidth}{!}{%
    \begin{tabular}{>{\ttfamily\small}p{\columnwidth}}
        \toprule
        \multicolumn{1}{l}{\textbf{Performance Counters/Events}} \\
        \midrule
        
        \rowcolor{gray!15}
        \multicolumn{1}{c}{\textbf{On-core Events}}\\
        Core Clock Cycles, Reference Clock Cycles,
        Temperature, Instructions (\textmu{}ops for Intel) issued, 
        Instructions (\textmu{}ops for Intel) retired,
        Un-utilized slots due to miss-speculation\\
        
        \rowcolor{gray!15}
        \multicolumn{1}{c}{\textbf{Un-core \& Memory Controller Events (per socket)}}\\
        \#Read/Write requests to DRAM (from all channels),
        \#Local DRAM accesses, \#Remote DRAM Accesses,
        \#Read/Write requests to DRAM (from all channels) from IO sources,
        \#PCIe Read, \#PCIe Write, QPI(for Intel)/Nest(for IBM) Transactions\\
        
        \rowcolor{gray!15}
        \multicolumn{1}{c}{\textbf{OS/Driver Events}}\\
        Free memory (CPU, GPU, FPGA),  Total memory (CPU, GPU, FPGA)\\
        \bottomrule
    \end{tabular}%
    }
\end{table}

\textbf{Implementation Details.}
We collect system-wide (for all processes) performance counter measurements for a variety of hardware events (described in \cref{tab:pmu_counters}).
The system wide collection leads to occasional spurious measurements (e.g., from interrupt handlers), however, this allows us to make holistic measurements (e.g., capture system calls or drivers that perform memory and DMA operations).
We make the minimum measurements to infer if a kernel scheduled to a CPU-hardware thread is core-bound (floating point- and integer-intensive).
This allows us to make scheduling decisions on co-located kernels, i.e., those that get scheduled to SMT/hyperthreads bound to a core.
The majority of measurements are made at the level of un-core events that captures performance of the memory interconnect and the system bus: to identify kernels that are bandwidth bottle necked.
We do not explicitly model GPU performance counters as low-level scheduling decisions (e.g., warp-level scheduling) in GPUs are obfuscated by the NVIDIA runtime/driver.

\textbf{NN Model.}
The second part of the POMDP-based scheduling model uses an NN (see \cref{fig:scheduler}) to learn the optimal policy with which to schedule user tasks given a belief state.
The NN takes two graphs as inputs. 
The first input is the belief state $\hat b_t$, encoded as vertex labels on a graph that describes the topology of a computer system (i.e., the organization shown in \cref{fig:system_topology}), and input labels that correspond to the locations of inputs in the topology. The color coding in \cref{fig:system_topology,fig:bn} shows a mapping (i.e., vertex labels) between nodes in the topology graph and $\hat b_t$.
The second input is the user's program expressed as a DFG.
We use \emph{graph network} (GN) layers~\cite{Battaglia2018} to ``embed'' the graphs into a set of \emph{embedding vectors}.
GNs have been shown to capture node, edge, and locality information.
We chose small, fully connected NNs for modeling the functional transformations in the GN layers.
Prior work in scheduling (e.g.,~\citet{Grandl2016, Wu2012}) has shown the benefit of considering temporal information to capture the dependencies of system resources over time as well as the time evolution of the user DFG.
We capture those relationships (between the embeddings of the input graphs) by using an RNN, specifically an LSTM layer~\cite{Hochreiter1997}.

The action space $\mathcal{A}$ of the model is fixed as the number of kernels/processors available in the system and is known ahead of time.
The action space consists of the following types of actions.
\begin{enumerate*}[label=(\roman*)]
    \item \emph{Execution actions} correspond to execution of a kernel on a processor/accelerator.
    \item \emph{Reconfiguration actions} correspond to reconfiguration of a single FPGA context to a kernel.
    \item \emph{No-Op actions} correspond to not scheduling any task in a particular scheduler  invocation. 
\end{enumerate*}
No-Ops are useful when the system resources are maximally subscribed, and execution of more tasks will hinder performance.
The scheduler is invoked every time there is an idle processor/accelerator in the system (i.e., every time a processor finishes the work assigned to it), causing the system to take one of the above actions.

\textbf{Reward Function.}
The reward $r_t$ is based on the objective of minimizing the runtime of a user DFG.
At time $t$, $r_t = - \sum_{i=0}^{t} \nicefrac{1}{T_i}$, where $T_i$ is the wall clock time taken to execute the $i$ actions executing in the system at time $t$. 
We picked $r_t$ as it represents the ``makespan'' of the schedule, a metric that is popularly used in the traditional scheduling literature and accurately represents the user-facing performance of the system.
Note that parallel actions are not double-counted in this formulation. 
The BN and NN models are trained end-to-end using minimization of \cref{eqn:loss_funcs} through back-propagation, as described in \cref{sec:model}. 

Implementation details of the BN and NN models are presented in the supplementary material.

    \section{Evaluation \& Discussion} \label{sec:results}
We evaluated the Symphony along the following dimensions.
\begin{enumerate*}[label=(\roman*)]
    \item \emph{How well does Symphony perform compared to the state of the art?}
    \item \emph{How does the Symphony's runtime affect scheduling decisions?}
    \item \emph{What are the savings in training time compared to traditional methods?} 
\end{enumerate*}
The evaluation testbed consisted of a rack-scale cluster of twelve IBM Power8 CPUs, two NVIDIA K40, six K80 GPUs, and two FPGAs.
We illustrated the generality of techniques on a variety of real-world workloads that used CPUs, GPUs, and FPGAs:
\begin{enumerate*}[label=(\roman*)]
    \item \emph{variant calling and genotyping analysis}~\cite{VanDerAuwera2013} on human genome datasets using tools presented in~\citet{Banerjee2016, Banerjee2017_FPL, Banerjee2018_ASAP, Li2009BWA, Li2010, Langmead2009, McKenna2010, Nothaft2015, Nothaft2015ms, Rimmer2014, Zaharia2011snap};
	\item \emph{epilepsy detection and localization}~\cite{Varatharajah2017} on intra-cranial electroencephalography data; and
	\item in online \emph{security analytics}~\cite{Cao2015} for intrusion detection systems.
\end{enumerate*}

\textbf{State of the Art.}
Traditional dynamic scheduling techniques (e.g.,~\citet{Isard2009, Giceva2014, Lyerly2018, Ousterhout2013, Zhuravlev2010, Zaharia2010}) use manually tuned heuristics (e.g., fairness, shortest-job-first) that prioritize simplicity and generality over achieving the best-case workload performance, often allocating coarse-grained resources (e.g., GBs of memory, CPU threads) and making simplifying assumptions about the underlying workload.
Several ML-based scheduling strategies have also been proposed, wherein the above heuristics are learned from data.
They use a variety of black-box ML models, e.g., model-free deep RL in~\cite{Mao2016, Mao2018}, collaborative filtering~\cite{Delimitrou2013, Delimitrou2014}, and other traditional ML techniques like SVMs (e.g.,~\citet{Mars2011, Mars2013, Yang2013, Zhang2014}).
A common theme in these techniques is that of treating the system as a black-box and performing scheduling to optimize application throughput metrics.
The above approaches are not well-suited to heterogeneous, accelerator-rich systems in which architectural diversity necessitates the use of low-level resources, which cannot be measured directly and are not semantically comparable across processors.
As points of comparison to Symphony, we used \textit{Graphene}~\cite{Grandl2016}, a heuristic-accelerated job shop optimization solver\footnote{Graphene was not originally designed to execute on heterogeneous systems. In the supplementary material, we explain modifications we made to the algorithm.}; \textit{Sparrow}~\cite{Ousterhout2013}, a randomized scheduler; and \textit{Paragon}~\cite{Delimitrou2013}, a collaborative filtering-based scheduler.

\textbf{Baseline for Comparison.}
We defined the \emph{oracle schedule} to correspond to the best performance possible for running an application on the evaluation system.
It corresponds to a completely isolated execution of an application.
Here, different concurrently executing kernels of the same application contend for resources and might cause performance degradation.
For the benchmark applications, we accounted for that by exhaustively executing schedules of the application DFGs to find the one with the lowest runtime (i.e., the \textit{oracle run}).
We measured the runtime of kernel $i$ in workload (in the oracle run) $j$ as $t_{i,j}^{\text{oracle}}$ across all kernels and workloads.
$t_{i,j}^{\text{oracle}}$ serves as the baseline for assessing the performance of Symphony.

\begin{figure*}[!t]
    \centering
    \begin{minipage}{.49\textwidth}
        \centering
        \includegraphics{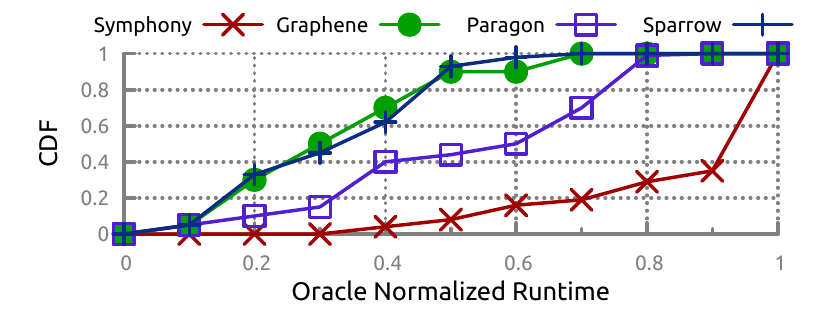}
        \vspace{-0.8cm}
        \caption{Comparing performance of Symphony to that of other popular schedulers for kernel executions in DFGs.}
        \label{fig:sched_comp}
    \end{minipage}%
    \hfill
    \begin{minipage}{.49\textwidth}
        \centering
        \includegraphics{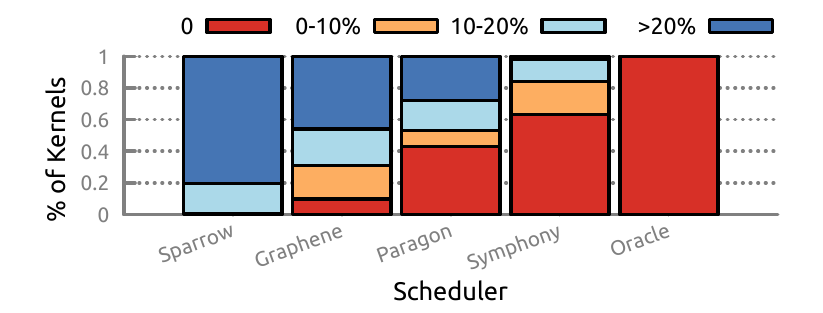}
        \vspace{-0.8cm}
        \caption{Percentage of application executions that show a degradation in performance.}
        \label{fig:decision_quality}
    \end{minipage}%
\end{figure*}

\textbf{Effectiveness of Scheduling Model.}
First, we quantified how well Symphony can handle scheduling of kernels in a DFG taking into account of resource contention and interference at
\begin{enumerate*}[label=(\roman*)]
    \item intra-DFG level; and 
    \item when executing with an unknown co-located workload utilizing compute and I/O resources.
\end{enumerate*} 
To do so, we measured the runtimes of each of the kernels $i$ in the workload $j$ (as above) to compute $t_{i,j}^{s}$ for each scheduler $s$ under test.
In \cref{fig:sched_comp}, we illustrate the distribution of oracle-normalized runtimes for each of the kernels in the workloads we tested, i.e., a distribution of $\nicefrac{t_{i,j}^{s}}{t_{i,j}^{\text{oracle}}}$ across 500 executions of the three above workloads.
In the figure, a distribution whose probability mass is closest to 1 is preferred, as it implies the least slowdown compared to the oracle.
We observe that the proposed technique significantly outperformed the state-of-the-art.
In our experiments, the median and tail (i.e., 99\textsuperscript{th} percentile) runtime of Symphony outperformed the second best (in this case, Paragon) by close to 32\%.
At the 99\textsuperscript{th} percentile, the generated schedules performed at a 6\% loss relative to the oracle.
Next, we quantified the performance of end-to-end user workloads, shown in \cref{fig:decision_quality}.
Here, we calculated $1 - \nicefrac{(\sum_it_{i,j}^{s})}{(\sum_it_{i,j}^{\text{oracle}})}$ for all 500 runs of the DFGs and grouped them into buckets of different kinds of normalized performance.
Symphony significantly outperformed the other scheduling techniques, running up to 60\% of the applications with no performance loss relative to the oracle execution, and the rest with a performance loss of less than 20\%.

\begin{figure*}[!t]
    \centering
    \begin{minipage}{.32\textwidth}
        \centering
        \includegraphics{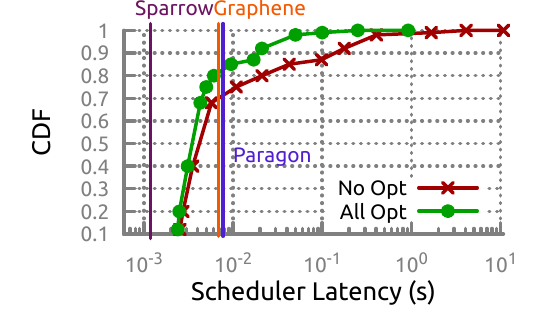}
        \vspace{-0.8cm}
        \caption{Symphony's latency (``All Opt'' \& ``No Opt'') compared to prior work.}
        \label{fig:sched_lat}
    \end{minipage}%
    \hfill
    \begin{minipage}{.32\textwidth}
        \centering
        \includegraphics{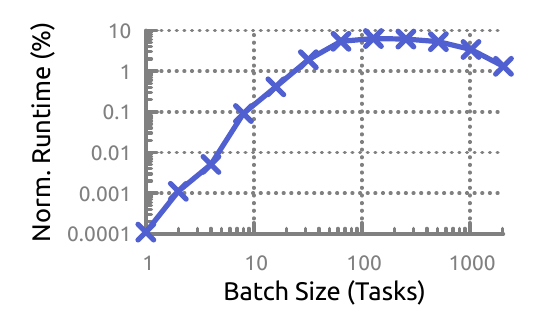}
        \vspace{-0.8cm}
        \caption{Symphony's performance (oracle normalized, in \%) with varying batch size.}
        \label{fig:batching}
    \end{minipage}%
    \hfill
    \begin{minipage}{.32\textwidth}
        \centering
        \includegraphics{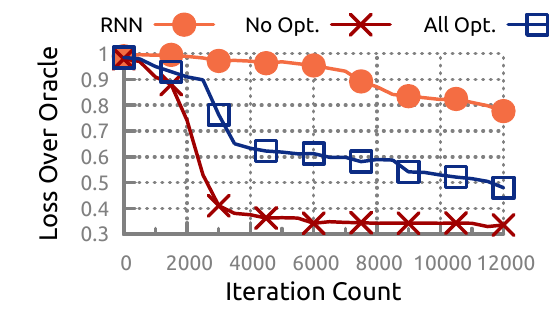}
        \vspace{-0.8cm}
        \caption{Training time for Symphony. An iteration is 2 RL episodes of 20 steps.}
        \label{fig:train_time}
    \end{minipage}%
\end{figure*}

\textbf{Latency.} 
There are two latencies to consider in comparing schedulers: the latency of the entire user workload (``LW'', shown in \cref{fig:sched_comp}), and the latency of the scheduler execution (``LS'', shown in \cref{fig:sched_lat}).
In \cref{fig:sched_lat}, we show two configurations of the Symphony scheduler:
\begin{enumerate*}[label=(\roman*)]
    \item ``No-Opt'' which uses a belief propagation-based update for the BN (and MCMC-based inference); and
    \item ``All-Opt'' which uses the sampling technique described in \cref{sec:model}, accelerators\footnote{The accelerators include an NVIDIA K80 GPU for NN inference and an FPGA for BN inference using \citet{Banerjee2019}.} to perform inference, and \emph{task batching} (described below).
\end{enumerate*}
LW ($\geq$ LS) is the user-facing metric of interest.
Symphony outperforms all baselines in terms of LW.
In terms of median LS, the Symphony is 1.8\texttimes{} and 1.6\texttimes{} faster than Paragon and Graphene, respectively.
In contrast, Sparrow, which randomly assigns tasks to processors, has $3.6\times$ lower median latency than Symphony.
However, the reduced LS comes at the cost of increased LW (see \cref{fig:sched_comp}).

\textbf{Batching Task Execution.}
A key concern with Symphony is its large tail latency (100\texttimes{} larger than its median; see \cref{fig:sched_lat}) compared to the other schedulers (which have deterministic runtime).
This increased latency is brought about by Symphony having to perform significantly more compute if the RL-policy-update is triggered.
The scheduler latency adversely affects LW as the time spent executing scheduler calls, is time not utilized to make progress on the user workload.
In order to deal with this issue, our evaluation executed Symphony on batches of tasks instead of single tasks, thereby amortizing the cost of executing Symphony across the batch.
Task batching works synergistically with the sampling based gradient propagation technique to reduce the tail latency by as much 12\texttimes{} (see \cref{fig:scheduler}).
\cref{fig:batching} demonstrates the average improvement in LW normalized to the oracle over a range of batch sizes.
We observe that the optimal value for batch size is about 128 tasks per batch.
This corresponds to the ``All Opt'' configuration in \cref{fig:sched_lat,fig:train_time} as well as \cref{fig:sched_comp,fig:decision_quality}. The ``No Opt'' configuration in \cref{fig:sched_lat} is computed at a batch size of one.

\textbf{Training Time.}
Finally, we quantified the improvement in training time offered by Symphony using the sampling-based gradient computation methodology presented in \cref{sec:model}.
We used the following baselines for evaluation: 
\begin{enumerate*}[label=(\roman*)]
    \item model-free RNN (labeled ``RNN'' in \cref{fig:train_time}); and
    \item the ``All Opt.'' and ``No Opt.'' configurations from above.
\end{enumerate*}
The RNN model here replaces the BN (and inference) and system-topology-embedding GN (in \cref{fig:scheduler}) with a 3-layer, fully connected NN to compute an embedding for $o_t$.
\cref{fig:train_time} illustrates the differences in performance of the these configurations with respect to degradation in performance of the user DFGs relative to the oracle schedule (i.e., $1 - \nicefrac{(\sum_it_{i,j}^{s})}{(\sum_it_{i,j}^{\text{oracle}})}$).
We observe that the RNN is significantly less sample-efficient than the proposed POMDP is; specifically, it is \textasciitilde{}2.2\texttimes{} worse than Symphony.
Further linearly extrapolating time to convergence from iteration $12\times10^3$, the RNN would need $> 48\times 10^3$ iterations to achieve the same accuracy as Symphony.

The difference in training time for the ``No Opt.'' and ``All Opt'' in \cref{fig:train_time} can be attributed to 
\begin{enumerate*}[label=(\roman*)]
    \item time taken to perform back-propagation for policy updates; and
    \item effective scheduler latency.
\end{enumerate*}
Linearly extrapolating the training-loss, we observe that ``All Opt'' is at least 4.3\texttimes{} more sample efficient than ``No Opt'' to reach a 30\% mean loss relative to the oracle.
That reduction is significant because the continuous churn of user workloads and machine configurations in a cloud, as pointed out in \citet{Mars2011}, would require that the scheduling model be periodically retrained.
In absolute terms, the ``All Opt'' configuration is able to achieve \textasciitilde{}30\% mean loss relative to the oracle scheduler in 700 hours of training and \textasciitilde{}4400 iterations of workload execution.
That corresponds to approximately 500 hours of system execution; hence, the total process takes 1200 hours.
Though this might appear to be over 7 weeks of time, in wall clock time this is approximately 2 week because we use parallel A3C-based training.
In fact, the limiting factor here is the availability of FPGAs, of which we have only 2 in the evaluation cluster, hence limiting the number of RL episodes that can be run in parallel.

    \section{Conclusion} \label{sec:conclusion}
This paper presents
\begin{enumerate*}[label=(\roman*)]
    \item a domain-driven Bayesian RL model for scheduling that captures the statistical dependencies between architectural resources; and
    \item a sampling-based technique that allows the computation of gradients of a Bayesian model without performing full probabilistic inference.
\end{enumerate*}
As data center architectures become more complex~\cite{Asanovic2014, Shao2015}, techniques like the one proposed here will be critical in the deployment of future accelerated applications.

\section*{Acknowledgments}
We thank K. Saboo, S. Lumetta, W-M. Hwu, K. Atchley, and J. Applequist for their insightful comments on the early drafts of this manuscript.
This research was supported in part by the National Science Foundation (NSF)
under Grant Nos. CNS 13-37732 and CNS 16-24790; by IBM under a Faculty Award and through equipment donations; and by Xilinx and Intel through equipment donations.
Any opinions, findings, and conclusions or recommendations expressed in this material are those of the authors and do not necessarily reflect the views of the NSF, IBM, Xilinx or Intel.

    \bibliographystyle{icml2020}
    \bibliography{references}

\begin{thebibliography}{79}
\providecommand{\natexlab}[1]{#1}
\providecommand{\url}[1]{\texttt{#1}}
\expandafter\ifx\csname urlstyle\endcsname\relax
  \providecommand{\doi}[1]{doi: #1}\else
  \providecommand{\doi}{doi: \begingroup \urlstyle{rm}\Url}\fi

\bibitem[Abadi et~al.(2016)Abadi, Barham, Chen, Chen, Davis, Dean, Devin,
  Ghemawat, Irving, Isard, Kudlur, Levenberg, Monga, Moore, Murray, Steiner,
  Tucker, Vasudevan, Warden, Wicke, Yu, and Zheng]{Abadi2016}
Abadi, M., Barham, P., Chen, J., Chen, Z., Davis, A., Dean, J., Devin, M.,
  Ghemawat, S., Irving, G., Isard, M., Kudlur, M., Levenberg, J., Monga, R.,
  Moore, S., Murray, D.~G., Steiner, B., Tucker, P., Vasudevan, V., Warden, P.,
  Wicke, M., Yu, Y., and Zheng, X.
\newblock {TensorFlow: A System for Large-scale Machine Learning}.
\newblock In \emph{Proceedings of the 12th USENIX Conference on Operating
  Systems Design and Implementation}, pp.\  265--283. USENIX Association, 2016.

\bibitem[Asanovi{\'c}(2014)]{Asanovic2014}
Asanovi{\'c}, K.
\newblock {FireBox: A Hardware Building Block for 2020 Warehouse-Scale
  Computers}.
\newblock Santa Clara, CA, February 2014. {USENIX} Association.

\bibitem[Asanovi{\'c} et~al.(2009)Asanovi{\'c}, Bodik, Demmel, Keaveny,
  Keutzer, Kubiatowicz, Morgan, Patterson, Sen, Wawrzynek, Wessel, and
  Yelick]{Asanovic2009}
Asanovi{\'c}, K., Bodik, R., Demmel, J., Keaveny, T., Keutzer, K., Kubiatowicz,
  J., Morgan, N., Patterson, D., Sen, K., Wawrzynek, J., Wessel, D., and
  Yelick, K.
\newblock {A View of the Parallel Computing Landscape}.
\newblock \emph{Commun. ACM}, 52\penalty0 (10):\penalty0 56--67, October 2009.

\bibitem[Astrom(1965)]{Astrom1965}
Astrom, K.~J.
\newblock Optimal control of {M}arkov processes with incomplete state
  information.
\newblock \emph{{Journal of Mathematical Analysis and Applications}},
  10\penalty0 (1):\penalty0 174--205, 1965.

\bibitem[{Banerjee} et~al.(2016){Banerjee}, Athreya, Mainzer, Jongeneel, Hwu,
  Kalbarczyk, and Iyer]{Banerjee2016}
{Banerjee}, S.~S., Athreya, A.~P., Mainzer, L.~S., Jongeneel, C.~V., Hwu,
  W.-M., Kalbarczyk, Z.~T., and Iyer, R.~K.
\newblock Efficient and scalable workflows for genomic analyses.
\newblock In \emph{Proceedings of the ACM International Workshop on
  Data-Intensive Distributed Computing}, DIDC '16, pp.\  27--36, 2016.

\bibitem[{Banerjee} et~al.(2017){Banerjee}, El{-}Hadedy, Tan, Kalbarczyk,
  Lumetta, and Iyer]{Banerjee2017_FPL}
{Banerjee}, S.~S., El{-}Hadedy, M., Tan, C.~Y., Kalbarczyk, Z.~T., Lumetta,
  S.~S., and Iyer, R.~K.
\newblock {On accelerating pair-HMM computations in programmable hardware}.
\newblock In \emph{Proc. 27th International Conference on Field Programmable
  Logic and Applications, {FPL} 2017, Ghent, Belgium, September 4-8, 2017},
  pp.\  1--8, 2017.

\bibitem[{Banerjee} et~al.(2019{\natexlab{a}}){Banerjee}, {El-Hadedy}, {Lim},
  {Kalbarczyk}, {Chen}, {Lumetta}, and {Iyer}]{Banerjee2018_ASAP}
{Banerjee}, S.~S., {El-Hadedy}, M., {Lim}, J.~B., {Kalbarczyk}, Z.~T., {Chen},
  D., {Lumetta}, S.~S., and {Iyer}, R.~K.
\newblock {ASAP: Accelerated Short-Read Alignment on Programmable Hardware}.
\newblock \emph{IEEE Transactions on Computers}, 68\penalty0 (3):\penalty0
  331--346, March 2019{\natexlab{a}}.

\bibitem[{Banerjee} et~al.(2019{\natexlab{b}}){Banerjee}, Kalbarczyk, and
  Iyer]{Banerjee2019}
{Banerjee}, S.~S., Kalbarczyk, Z.~T., and Iyer, R.~K.
\newblock {AcMC\textsuperscript{2} : Accelerating Markov Chain Monte Carlo
  Algorithms for Probabilistic Models}.
\newblock In \emph{{Proceedings of the Twenty-Fourth International Conference
  on Architectural Support for Programming Languages and Operating Systems}},
  pp.\  515--528, 2019{\natexlab{b}}.

\bibitem[Battaglia et~al.(2018)Battaglia, Hamrick, Bapst, Sanchez-Gonzalez,
  Zambaldi, Malinowski, Tacchetti, Raposo, Santoro, Faulkner,
  et~al.]{Battaglia2018}
Battaglia, P.~W., Hamrick, J.~B., Bapst, V., Sanchez-Gonzalez, A., Zambaldi,
  V., Malinowski, M., Tacchetti, A., Raposo, D., Santoro, A., Faulkner, R.,
  et~al.
\newblock Relational inductive biases, deep learning, and graph networks.
\newblock \emph{arXiv preprint arXiv:1806.01261}, 2018.

\bibitem[Binder et~al.(1997)Binder, Koller, Russell, and Kanazawa]{Binder1997}
Binder, J., Koller, D., Russell, S., and Kanazawa, K.
\newblock {Adaptive Probabilistic Networks with Hidden Variables}.
\newblock \emph{Machine Learning}, 29\penalty0 (2/3):\penalty0 213--244, 1997.

\bibitem[Broquedis et~al.(2010)Broquedis, Clet-Ortega, Moreaud, Furmento,
  Goglin, Mercier, Thibault, and Namyst]{Broquedis2010}
Broquedis, F., Clet-Ortega, J., Moreaud, S., Furmento, N., Goglin, B., Mercier,
  G., Thibault, S., and Namyst, R.
\newblock {hwloc: A Generic Framework for Managing Hardware Affinities in HPC
  Applications}.
\newblock In \emph{Proc. 2010 18th Euromicro Conference on Parallel,
  Distributed and Network-based Processing}, pp.\  180--186, Feb 2010.

\bibitem[Cao et~al.(2015)Cao, Badger, Kalbarczyk, Iyer, and Slagell]{Cao2015}
Cao, P., Badger, E., Kalbarczyk, Z., Iyer, R., and Slagell, A.
\newblock Preemptive intrusion detection: Theoretical framework and real-world
  measurements.
\newblock In \emph{Proceedings of the 2015 Symposium and Bootcamp on the
  Science of Security}, HotSoS '15, pp.\  5:1--5:12, 2015.

\bibitem[Chambers et~al.(2010)Chambers, Raniwala, Perry, Adams, Henry,
  Bradshaw, and Nathan]{Chambers2010}
Chambers, C., Raniwala, A., Perry, F., Adams, S., Henry, R., Bradshaw, R., and
  Nathan.
\newblock {FlumeJava: Easy, Efficient Data-Parallel Pipelines}.
\newblock In \emph{ACM SIGPLAN Conference on Programming Language Design and
  Implementation (PLDI)}, pp.\  363--375, 2010.

\bibitem[Chowdhury et~al.(2014)Chowdhury, Zhong, and Stoica]{Chowdhury2014}
Chowdhury, M., Zhong, Y., and Stoica, I.
\newblock Efficient coflow scheduling with varys.
\newblock In \emph{Proceedings of the 2014 ACM Conference on SIGCOMM}, SIGCOMM
  '14, pp.\  443--454, 2014.

\bibitem[Dagum \& Luby(1993)Dagum and Luby]{Dagum1993}
Dagum, P. and Luby, M.
\newblock {Approximating probabilistic inference in Bayesian belief networks is
  NP-hard}.
\newblock \emph{{Artificial Intelligence}}, 60\penalty0 (1):\penalty0 141--153,
  1993.

\bibitem[Delimitrou \& Kozyrakis(2013)Delimitrou and Kozyrakis]{Delimitrou2013}
Delimitrou, C. and Kozyrakis, C.
\newblock {Paragon: QoS-aware Scheduling for Heterogeneous Datacenters}.
\newblock \emph{SIGPLAN Not.}, 48\penalty0 (4):\penalty0 77--88, March 2013.

\bibitem[Delimitrou \& Kozyrakis(2014)Delimitrou and Kozyrakis]{Delimitrou2014}
Delimitrou, C. and Kozyrakis, C.
\newblock {Quasar: Resource-efficient and QoS-aware Cluster Management}.
\newblock In \emph{Proceedings of the 19th International Conference on
  Architectural Support for Programming Languages and Operating Systems},
  ASPLOS '14, pp.\  127--144, 2014.

\bibitem[Dhariwal et~al.(2017)Dhariwal, Hesse, Klimov, Nichol, Plappert,
  Radford, Schulman, Sidor, Wu, and Zhokhov]{Dhariwal2017}
Dhariwal, P., Hesse, C., Klimov, O., Nichol, A., Plappert, M., Radford, A.,
  Schulman, J., Sidor, S., Wu, Y., and Zhokhov, P.
\newblock {OpenAI Baselines}.
\newblock \url{https://github.com/openai/baselines}, 2017.

\bibitem[Doweck(2016)]{IntelHotChips}
Doweck, J.
\newblock {Inside 6th generation Intel Core code named Skylake:: New
  Microarchitecture and Power Management}.
\newblock
  \url{https://www.hotchips.org/wp-content/uploads/hc_archives/hc28/HC28.23-Tuesday-Epub/HC28.23.90-High-Perform-Epub/HC28.23.911-Skylake-Doweck-Intel_SK3-r13b.pdf},
  2016.
\newblock Accessed 2019-03-05.

\bibitem[Dreyer \& Alpert(1997)Dreyer and Alpert]{Dreyer1997}
Dreyer, R.~S. and Alpert, D.~B.
\newblock Apparatus for monitoring the performance of a microprocessor, August
  1997.
\newblock US Patent 5,657,253.

\bibitem[Foerster et~al.(2016)Foerster, Assael, de~Freitas, and
  Whiteson]{Foerster2016}
Foerster, J.~N., Assael, Y.~M., de~Freitas, N., and Whiteson, S.
\newblock {Learning to communicate to solve riddles with deep distributed
  recurrent Q-networks}.
\newblock \emph{arXiv preprint arXiv:1602.02672}, 2016.

\bibitem[Gelman et~al.(1995)Gelman, Carlin, Stern, and Rubin]{Gelman1995}
Gelman, A., Carlin, J., Stern, H., and Rubin, D.
\newblock \emph{Bayesian Data Analysis}.
\newblock Chapman \& Hall, New York, 1995.

\bibitem[Giceva et~al.(2014)Giceva, Alonso, Roscoe, and Harris]{Giceva2014}
Giceva, J., Alonso, G., Roscoe, T., and Harris, T.
\newblock Deployment of query plans on multicores.
\newblock \emph{Proc. VLDB Endow.}, 8\penalty0 (3):\penalty0 233--244, November
  2014.

\bibitem[Grandl et~al.(2016{\natexlab{a}})Grandl, Chowdhury, Akella, and
  Ananthanarayanan]{Grandl2016:2}
Grandl, R., Chowdhury, M., Akella, A., and Ananthanarayanan, G.
\newblock {Altruistic Scheduling in Multi-resource Clusters}.
\newblock In \emph{Proceedings of the 12th USENIX Conference on Operating
  Systems Design and Implementation}, pp.\  65--80. USENIX Association,
  2016{\natexlab{a}}.

\bibitem[Grandl et~al.(2016{\natexlab{b}})Grandl, Kandula, Rao, Akella, and
  Kulkarni]{Grandl2016}
Grandl, R., Kandula, S., Rao, S., Akella, A., and Kulkarni, J.
\newblock {Graphene: Packing and Dependency-aware Scheduling for Data-parallel
  Clusters}.
\newblock In \emph{Proceedings of the 12th USENIX Conference on Operating
  Systems Design and Implementation}, pp.\  81--97, 2016{\natexlab{b}}.

\bibitem[Hall et~al.(2017)Hall, Bergner, Housfater, Kandasamy, Magno, Mericas,
  Munroe, Oliveira, Schmidt, Schmidt, et~al.]{IBM2017_Perf}
Hall, B., Bergner, P., Housfater, A.~S., Kandasamy, M., Magno, T., Mericas, A.,
  Munroe, S., Oliveira, M., Schmidt, B., Schmidt, W., et~al.
\newblock \emph{Performance optimization and tuning techniques for IBM Power
  Systems processors including IBM POWER8}.
\newblock IBM Redbooks, 2017.

\bibitem[Hausknecht \& Stone(2015)Hausknecht and Stone]{Hausknecht2015}
Hausknecht, M. and Stone, P.
\newblock {Deep recurrent Q-learning for partially observable MDPs}.
\newblock In \emph{2015 AAAI Fall Symposium Series}, 2015.

\bibitem[Hindman et~al.(2011)Hindman, Konwinski, Zaharia, Ghodsi, Joseph, Katz,
  Shenker, and Stoica]{Hindman2011}
Hindman, B., Konwinski, A., Zaharia, M., Ghodsi, A., Joseph, A.~D., Katz, R.,
  Shenker, S., and Stoica, I.
\newblock Mesos: A platform for fine-grained resource sharing in the data
  center.
\newblock In \emph{Proceedings of the 8th USENIX Conference on Networked
  Systems Design and Implementation}, NSDI'11, pp.\  295--308. USENIX
  Association, 2011.

\bibitem[Hochreiter \& Schmidhuber(1997)Hochreiter and
  Schmidhuber]{Hochreiter1997}
Hochreiter, S. and Schmidhuber, J.
\newblock Long short-term memory.
\newblock \emph{{Neural computation}}, 9\penalty0 (8):\penalty0 1735--1780,
  1997.

\bibitem[Igl et~al.(2018)Igl, Zintgraf, Le, Wood, and Whiteson]{Igl2018}
Igl, M., Zintgraf, L., Le, T.~A., Wood, F., and Whiteson, S.
\newblock {Deep variational reinforcement learning for POMDPs}.
\newblock \emph{arXiv preprint arXiv:1806.02426}, 2018.

\bibitem[{Intel Corp.}(2014)]{IntelAORM}
{Intel Corp.}
\newblock Intel 64 and ia-32 architectures optimization reference manual.
\newblock \emph{Intel Corporation, Sept}, 2014.

\bibitem[{Intel Corp.}(2016)]{IntelSDM}
{Intel Corp.}
\newblock {Intel® 64 and IA-32 Architectures Software Developer Manuals}.
\newblock \url{https://software.intel.com/en-us/articles/intel-sdm}, 2016.
\newblock Accessed 2019-03-05.

\bibitem[Isard et~al.(2009)Isard, Prabhakaran, Currey, Wieder, Talwar, and
  Goldberg]{Isard2009}
Isard, M., Prabhakaran, V., Currey, J., Wieder, U., Talwar, K., and Goldberg,
  A.
\newblock Quincy: Fair scheduling for distributed computing clusters.
\newblock In \emph{Proceedings of the ACM SIGOPS 22nd Symposium on Operating
  Systems Principles}, SOSP '09, pp.\  261--276, 2009.

\bibitem[Jaderberg et~al.(2016)Jaderberg, Mnih, Czarnecki, Schaul, Leibo,
  Silver, and Kavukcuoglu]{Jaderberg2016}
Jaderberg, M., Mnih, V., Czarnecki, W.~M., Schaul, T., Leibo, J.~Z., Silver,
  D., and Kavukcuoglu, K.
\newblock Reinforcement learning with unsupervised auxiliary tasks.
\newblock \emph{arXiv preprint arXiv:1611.05397}, 2016.

\bibitem[Kaelbling et~al.(1998)Kaelbling, Littman, and
  Cassandra]{Kaelbling1998}
Kaelbling, L.~P., Littman, M.~L., and Cassandra, A.~R.
\newblock Planning and acting in partially observable stochastic domains.
\newblock \emph{{Artificial Intelligence}}, 101\penalty0 (1--2):\penalty0
  99--134, 1998.

\bibitem[Karkus et~al.(2017)Karkus, Hsu, and Lee]{Karkus2017}
Karkus, P., Hsu, D., and Lee, W.~S.
\newblock {QMDP-Net: Deep learning for planning under partial observability}.
\newblock In \emph{{Advances in Neural Information Processing Systems}}, pp.\
  4694--4704, 2017.

\bibitem[Kleen(2010)]{PMUTools}
Kleen, A.
\newblock {PMU-Tools}.
\newblock \url{https://github.com/andikleen/pmu-tools}, 2010.
\newblock Accessed 2019-03-05.

\bibitem[Konda \& Tsitsiklis(2000)Konda and Tsitsiklis]{Konda2000}
Konda, V.~R. and Tsitsiklis, J.~N.
\newblock Actor-critic algorithms.
\newblock In \emph{Advances in neural information processing systems}, pp.\
  1008--1014, 2000.

\bibitem[Langmead et~al.(2009)Langmead, Trapnell, Pop, and
  Salzberg]{Langmead2009}
Langmead, B., Trapnell, C., Pop, M., and Salzberg, S.~L.
\newblock Ultrafast and memory-efficient alignment of short {DNA} sequences to
  the human genome.
\newblock \emph{Genome Biol}, 10\penalty0 (3):\penalty0 R25, 2009.

\bibitem[Li \& Durbin(2009)Li and Durbin]{Li2009BWA}
Li, H. and Durbin, R.
\newblock Fast and accurate short-read alignment with burrows-wheeler
  rransform.
\newblock \emph{Bioinformatics}, 25\penalty0 (14):\penalty0 1754--1760, may
  2009.
\newblock \doi{10.1093/bioinformatics/btp324}.
\newblock URL \url{http://dx.doi.org/10.1093/bioinformatics/btp324}.

\bibitem[Li \& Durbin(2010)Li and Durbin]{Li2010}
Li, H. and Durbin, R.
\newblock Fast and accurate long-read alignment with {B}urrows--{W}heeler
  transform.
\newblock \emph{Bioinformatics}, 26\penalty0 (5):\penalty0 589--595, 2010.

\bibitem[Li et~al.(2009)Li, Handsaker, Wysoker, Fennell, Ruan, Homer, Marth,
  Abecasis, Durbin, et~al.]{Li2009}
Li, H., Handsaker, B., Wysoker, A., Fennell, T., Ruan, J., Homer, N., Marth,
  G., Abecasis, G., Durbin, R., et~al.
\newblock The sequence alignment/map format and {SAM}tools.
\newblock \emph{Bioinformatics}, 25\penalty0 (16):\penalty0 2078--2079, 2009.

\bibitem[Lyerly et~al.(2018)Lyerly, Murray, Barbalace, and
  Ravindran]{Lyerly2018}
Lyerly, R., Murray, A., Barbalace, A., and Ravindran, B.
\newblock Aira: A framework for flexible compute kernel execution in
  heterogeneous platforms.
\newblock \emph{IEEE Transactions on Parallel and Distributed Systems},
  29\penalty0 (2):\penalty0 269--282, Feb 2018.
\newblock ISSN 1045-9219.
\newblock \doi{10.1109/TPDS.2017.2761748}.

\bibitem[Mao et~al.(2016)Mao, Alizadeh, Menache, and Kandula]{Mao2016}
Mao, H., Alizadeh, M., Menache, I., and Kandula, S.
\newblock Resource management with deep reinforcement learning.
\newblock In \emph{Proceedings of the 15th ACM Workshop on Hot Topics in
  Networks}, pp.\  50--56. ACM, 2016.

\bibitem[Mao et~al.(2018)Mao, Schwarzkopf, Venkatakrishnan, Meng, and
  Alizadeh]{Mao2018}
Mao, H., Schwarzkopf, M., Venkatakrishnan, S.~B., Meng, Z., and Alizadeh, M.
\newblock Learning scheduling algorithms for data processing clusters.
\newblock \emph{arXiv preprint arXiv:1810.01963}, 2018.

\bibitem[Mars \& Tang(2013)Mars and Tang]{Mars2013}
Mars, J. and Tang, L.
\newblock Whare-map: Heterogeneity in "homogeneous" warehouse-scale computers.
\newblock \emph{SIGARCH Comput. Archit. News}, 41\penalty0 (3):\penalty0
  619--630, June 2013.

\bibitem[Mars et~al.(2011)Mars, Tang, and Hundt]{Mars2011}
Mars, J., Tang, L., and Hundt, R.
\newblock Heterogeneity in ``homogeneous'' warehouse-scale computers: A
  performance opportunity.
\newblock \emph{IEEE Comput. Archit. Lett.}, 10\penalty0 (2):\penalty0 29--32,
  July 2011.
\newblock ISSN 1556-6056.

\bibitem[Mastrolilli \& Svensson(2008)Mastrolilli and
  Svensson]{Mastrolilli2008}
Mastrolilli, M. and Svensson, O.
\newblock (acyclic) job shops are hard to approximate.
\newblock In \emph{2008 49th Annual IEEE Symposium on Foundations of Computer
  Science}, pp.\  583--592, Oct 2008.
\newblock \doi{10.1109/FOCS.2008.36}.

\bibitem[Mastrolilli \& Svensson(2009)Mastrolilli and
  Svensson]{Mastrolilli2009}
Mastrolilli, M. and Svensson, O.
\newblock Improved bounds for flow shop scheduling.
\newblock In \emph{International Colloquium on Automata, Languages, and
  Programming}, pp.\  677--688. Springer, 2009.

\bibitem[McCool et~al.(2012)McCool, Reinders, and Robison]{McCool2012}
McCool, M., Reinders, J., and Robison, A.
\newblock \emph{Structured Parallel Programming: Patterns for Efficient
  Computation}.
\newblock Morgan Kaufmann Publishers Inc., San Francisco, CA, USA, 1st edition,
  2012.
\newblock ISBN 9780123914439, 9780124159938.

\bibitem[McKenna et~al.(2010)McKenna, Hanna, Banks, Sivachenko, Cibulskis,
  Kernytsky, Garimella, Altshuler, Gabriel, Daly, and DePristo]{McKenna2010}
McKenna, A., Hanna, M., Banks, E., Sivachenko, A., Cibulskis, K., Kernytsky,
  A., Garimella, K., Altshuler, D., Gabriel, S., Daly, M., and DePristo, M.~A.
\newblock The genome analysis toolkit: A {MapReduce} framework for analyzing
  next-generation {DNA} sequencing data.
\newblock \emph{Genome Research}, 20\penalty0 (9):\penalty0 1297--1303, jul
  2010.
\newblock \doi{10.1101/gr.107524.110}.

\bibitem[Mnih et~al.(2015)Mnih, Kavukcuoglu, Silver, Rusu, Veness, Bellemare,
  Graves, Riedmiller, Fidjeland, Ostrovski, et~al.]{Mnih2015}
Mnih, V., Kavukcuoglu, K., Silver, D., Rusu, A.~A., Veness, J., Bellemare,
  M.~G., Graves, A., Riedmiller, M., Fidjeland, A.~K., Ostrovski, G., et~al.
\newblock Human-level control through deep reinforcement learning.
\newblock \emph{Nature}, 518\penalty0 (7540):\penalty0 529, 2015.

\bibitem[Mnih et~al.(2016)Mnih, Badia, Mirza, Graves, Harley, Lillicrap,
  Silver, and Kavukcuoglu]{Mnih2016}
Mnih, V., Badia, A.~P., Mirza, M., Graves, A., Harley, T., Lillicrap, T.~P.,
  Silver, D., and Kavukcuoglu, K.
\newblock Asynchronous methods for deep reinforcement learning.
\newblock In \emph{Proceedings of the 33rd International Conference on
  International Conference on Machine Learning - Volume 48}, ICML'16, pp.\
  1928--1937. JMLR.org, 2016.

\bibitem[Narasimhan et~al.(2015)Narasimhan, Kulkarni, and
  Barzilay]{Narasimhan2015}
Narasimhan, K., Kulkarni, T., and Barzilay, R.
\newblock Language understanding for text-based games using deep reinforcement
  learning.
\newblock \emph{arXiv preprint arXiv:1506.08941}, 2015.

\bibitem[Nothaft(2015)]{Nothaft2015ms}
Nothaft, F.
\newblock Scalable genome resequencing with adam and avocado.
\newblock Master's thesis, EECS Department, University of California, Berkeley,
  May 2015.

\bibitem[Nothaft et~al.(2015)Nothaft, Massie, Danford, Zhang, Laserson,
  Yeksigian, Kottalam, Ahuja, Hammerbacher, Linderman, Franklin, Joseph, and
  Patterson]{Nothaft2015}
Nothaft, F., Massie, M., Danford, T., Zhang, Z., Laserson, U., Yeksigian, C.,
  Kottalam, J., Ahuja, A., Hammerbacher, J., Linderman, M., Franklin, M.~J.,
  Joseph, A.~D., and Patterson, D.~A.
\newblock Rethinking data-intensive science using scalable analytics systems.
\newblock In \emph{Proceedings of the 2015 ACM SIGMOD International Conference
  on Management of Data}, SIGMOD '15, pp.\  631--646, New York, NY, USA, 2015.
  ACM.
\newblock ISBN 978-1-4503-2758-9.
\newblock \doi{10.1145/2723372.2742787}.

\bibitem[Ousterhout et~al.(2013)Ousterhout, Wendell, Zaharia, and
  Stoica]{Ousterhout2013}
Ousterhout, K., Wendell, P., Zaharia, M., and Stoica, I.
\newblock Sparrow: Distributed, low latency scheduling.
\newblock In \emph{Proceedings of the Twenty-Fourth ACM Symposium on Operating
  Systems Principles}, SOSP '13, pp.\  69--84, New York, NY, USA, 2013. ACM.
\newblock ISBN 978-1-4503-2388-8.
\newblock \doi{10.1145/2517349.2522716}.

\bibitem[Rimmer et~al.(2014)Rimmer, Phan, Mathieson, Iqbal, Twigg, Wilkie,
  McVean, and Lunter]{Rimmer2014}
Rimmer, A., Phan, H., Mathieson, I., Iqbal, Z., Twigg, S. R.~F., Wilkie, A.
  O.~M., McVean, G., and Lunter, G.
\newblock Integrating mapping-, assembly- and haplotype-based approaches for
  calling variants in clinical sequencing applications.
\newblock \emph{Nature Genetics}, 46\penalty0 (8):\penalty0 912--918, jul 2014.

\bibitem[Russell et~al.(1995)Russell, Binder, Koller, and
  Kanazawa]{Russell1995}
Russell, S., Binder, J., Koller, D., and Kanazawa, K.
\newblock Local learning in probabilistic networks with hidden variables.
\newblock In \emph{Proceedings of the 14th International Joint Conference on
  Artificial Intelligence - Volume 2}, IJCAI'95, pp.\  1146--1152, San
  Francisco, CA, USA, 1995. Morgan Kaufmann Publishers Inc.
\newblock ISBN 1-55860-363-8.

\bibitem[Shachter(2013)]{Shachter2013}
Shachter, R.~D.
\newblock Bayes-ball: The rational pastime (for determining irrelevance and
  requisite information in belief networks and influence diagrams).
\newblock \emph{arXiv preprint arXiv:1301.7412}, 2013.

\bibitem[Shao \& Brooks(2015)Shao and Brooks]{Shao2015}
Shao, Y. and Brooks, D.
\newblock \emph{Research Infrastructures for Hardware Accelerators}.
\newblock Synthesis Lectures on Computer Architecture. Morgan \& Claypool
  Publishers, 2015.

\bibitem[Silver et~al.(2017)Silver, van Hasselt, Hessel, Schaul, Guez, Harley,
  Dulac-Arnold, Reichert, Rabinowitz, Barreto, et~al.]{Silver2017}
Silver, D., van Hasselt, H., Hessel, M., Schaul, T., Guez, A., Harley, T.,
  Dulac-Arnold, G., Reichert, D., Rabinowitz, N., Barreto, A., et~al.
\newblock {The Predictron: End-to-end learning and planning}.
\newblock In \emph{Proceedings of the 34th International Conference on Machine
  Learning-Volume 70}, pp.\  3191--3199. JMLR, 2017.

\bibitem[Stuecheli et~al.(2015)Stuecheli, Blaner, Johns, and Siegel]{CAPI2015}
Stuecheli, J., Blaner, B., Johns, C.~R., and Siegel, M.~S.
\newblock {CAPI: A Coherent Accelerator Processor Interface}.
\newblock \emph{IBM Journal of Research and Development}, 59\penalty0
  (1):\penalty0 7:1--7:7, Jan 2015.
\newblock ISSN 0018-8646.
\newblock \doi{10.1147/JRD.2014.2380198}.

\bibitem[Sudhakar \& Srinivasan(2019)Sudhakar and Srinivasan]{IBMDMAPerf}
Sudhakar, A.~T. and Srinivasan, M.
\newblock {IBM POWER in-memory collection counters}.
\newblock
  \url{https://developer.ibm.com/articles/power9-in-memory-collection-counters/},
  2019.
\newblock Accessed 2019-03-05.

\bibitem[Terpstra et~al.(2010)Terpstra, Jagode, You, and
  Dongarra]{Terpstra2010}
Terpstra, D., Jagode, H., You, H., and Dongarra, J.
\newblock {Collecting Performance Data with PAPI-C}.
\newblock In M{\"u}ller, M.~S., Resch, M.~M., Schulz, A., and Nagel, W.~E.
  (eds.), \emph{Tools for High Performance Computing 2009}, pp.\  157--173,
  Berlin, Heidelberg, 2010. Springer Berlin Heidelberg.

\bibitem[Van~der Auwera et~al.(2013)Van~der Auwera, Carneiro, Hartl, Poplin,
  del Angel, Levy-Moonshine, Jordan, Shakir, Roazen, Thibault, Banks,
  Garimella, Altshuler, Gabriel, and DePristo]{VanDerAuwera2013}
Van~der Auwera, G.~A., Carneiro, M.~O., Hartl, C., Poplin, R., del Angel, G.,
  Levy-Moonshine, A., Jordan, T., Shakir, K., Roazen, D., Thibault, J., Banks,
  E., Garimella, K.~V., Altshuler, D., Gabriel, S., and DePristo, M.~A.
\newblock From fastq data to high-confidence variant calls: The genome analysis
  toolkit best practices pipeline.
\newblock \emph{Current Protocols in Bioinformatics}, 43\penalty0 (1):\penalty0
  11.10.1--11.10.33, 2013.

\bibitem[Varatharajah et~al.(2017)Varatharajah, Chong, Saboo, Berry, Brinkmann,
  Worrell, and Iyer]{Varatharajah2017}
Varatharajah, Y., Chong, M.~J., Saboo, K., Berry, B., Brinkmann, B., Worrell,
  G., and Iyer, R.
\newblock {EEG-GRAPH: A Factor-Graph-Based Model for Capturing Spatial,
  Temporal, and Observational Relationships in Electroencephalograms}.
\newblock In \emph{Advances in Neural Information Processing Systems}, pp.\
  5377--5386, 2017.

\bibitem[Weaver \& McKee(2008)Weaver and McKee]{Weaver2008}
Weaver, V.~M. and McKee, S.~A.
\newblock Can hardware performance counters be trusted?
\newblock In \emph{2008 IEEE International Symposium on Workload
  Characterization}, pp.\  141--150. IEEE, 2008.

\bibitem[Wu et~al.(2012)Wu, Diamos, Cadambi, and Yalamanchili]{Wu2012}
Wu, H., Diamos, G., Cadambi, S., and Yalamanchili, S.
\newblock {Kernel Weaver: Automatically Fusing Database Primitives for
  Efficient GPU Computation}.
\newblock In \emph{2012 45th Annual IEEE/ACM International Symposium on
  Microarchitecture}, pp.\  107--118, Dec 2012.

\bibitem[{Xu} et~al.(2018){Xu}, {Butt}, {Lim}, and {Kannan}]{Xu2018}
{Xu}, L., {Butt}, A.~R., {Lim}, S., and {Kannan}, R.
\newblock {A Heterogeneity-Aware Task Scheduler for Spark}.
\newblock In \emph{Proc. 2018 IEEE International Conference on Cluster
  Computing (CLUSTER)}, pp.\  245--256, Sep. 2018.

\bibitem[Yang et~al.(2013)Yang, Breslow, Mars, and Tang]{Yang2013}
Yang, H., Breslow, A., Mars, J., and Tang, L.
\newblock {Bubble-flux: Precise Online QoS Management for Increased Utilization
  in Warehouse Scale Computers}.
\newblock In \emph{Proceedings of the 40th Annual International Symposium on
  Computer Architecture}, ISCA '13, pp.\  607--618, 2013.

\bibitem[Yasin(2014)]{Yasin2014}
Yasin, A.
\newblock {A Top-Down method for performance analysis and counters
  architecture}.
\newblock In \emph{Proc. 2014 IEEE International Symposium on Performance
  Analysis of Systems and Software (ISPASS)}, pp.\  35--44, March 2014.

\bibitem[Zaharia et~al.(2010)Zaharia, Borthakur, Sen~Sarma, Elmeleegy, Shenker,
  and Stoica]{Zaharia2010}
Zaharia, M., Borthakur, D., Sen~Sarma, J., Elmeleegy, K., Shenker, S., and
  Stoica, I.
\newblock {Delay Scheduling: A Simple Technique for Achieving Locality and
  Fairness in Cluster Scheduling}.
\newblock In \emph{Proceedings of the 5th European Conference on Computer
  Systems}, pp.\  265--278, 2010.

\bibitem[Zaharia et~al.(2011)Zaharia, Bolosky, Curtis, Fox, Patterson, Shenker,
  Stoica, Karp, and Sittler]{Zaharia2011snap}
Zaharia, M., Bolosky, W.~J., Curtis, K., Fox, A., Patterson, D., Shenker, S.,
  Stoica, I., Karp, R.~M., and Sittler, T.
\newblock Faster and more accurate sequence alignment with {SNAP}.
\newblock \emph{arXiv preprint arXiv:1111.5572}, 2011.

\bibitem[Zaharia et~al.(2012)Zaharia, Chowdhury, Das, Dave, Ma, McCauley,
  Franklin, Shenker, and Stoica]{Zaharia2012}
Zaharia, M., Chowdhury, M., Das, T., Dave, A., Ma, J., McCauley, M., Franklin,
  M.~J., Shenker, S., and Stoica, I.
\newblock {Resilient Distributed Datasets: A Fault-tolerant Abstraction for
  In-memory Cluster Computing}.
\newblock In \emph{Proceedings of the 9th USENIX Conference on Networked
  Systems Design and Implementation}, NSDI'12, pp.\  15--28, 2012.

\bibitem[Zhang et~al.(2014)Zhang, Laurenzano, Mars, and Tang]{Zhang2014}
Zhang, Y., Laurenzano, M.~A., Mars, J., and Tang, L.
\newblock {SMiTe: Precise QoS Prediction on Real-System SMT Processors to
  Improve Utilization in Warehouse Scale Computers}.
\newblock In \emph{2014 47th Annual IEEE/ACM International Symposium on
  Microarchitecture}, pp.\  406--418, Dec 2014.

\bibitem[Zhu et~al.(2018)Zhu, Li, Poupart, and Miao]{Zhu2018}
Zhu, P., Li, X., Poupart, P., and Miao, G.
\newblock {On improving deep reinforcement learning for POMDPs}.
\newblock \emph{arXiv preprint arXiv:1804.06309}, 2018.

\bibitem[Zhuravlev et~al.(2010)Zhuravlev, Blagodurov, and
  Fedorova]{Zhuravlev2010}
Zhuravlev, S., Blagodurov, S., and Fedorova, A.
\newblock {Addressing Shared Resource Contention in Multicore Processors via
  Scheduling}.
\newblock \emph{SIGPLAN Not.}, 45\penalty0 (3):\penalty0 129--142, March 2010.

\bibitem[Zook et~al.(2016)Zook, Catoe, McDaniel, Vang, Spies, Sidow, Weng, Liu,
  Mason, Alexander, Henaff, McIntyre, Chandramohan, Chen, Jaeger, Moshrefi,
  Pham, Stedman, Liang, Saghbini, Dzakula, Hastie, Cao, Deikus, Schadt, Sebra,
  Bashir, Truty, Chang, Gulbahce, Zhao, Ghosh, Hyland, Fu, Chaisson, Xiao,
  Trow, Sherry, Zaranek, Ball, Bobe, Estep, Church, Marks,
  Kyriazopoulou-Panagiotopoulou, Zheng, Schnall-Levin, Ordonez, Mudivarti,
  Giorda, Sheng, Rypdal, and Salit]{Zook2016}
Zook, J.~M., Catoe, D., McDaniel, J., Vang, L., Spies, N., Sidow, A., Weng, Z.,
  Liu, Y., Mason, C.~E., Alexander, N., Henaff, E., McIntyre, A.~B.,
  Chandramohan, D., Chen, F., Jaeger, E., Moshrefi, A., Pham, K., Stedman, W.,
  Liang, T., Saghbini, M., Dzakula, Z., Hastie, A., Cao, H., Deikus, G.,
  Schadt, E., Sebra, R., Bashir, A., Truty, R.~M., Chang, C.~C., Gulbahce, N.,
  Zhao, K., Ghosh, S., Hyland, F., Fu, Y., Chaisson, M., Xiao, C., Trow, J.,
  Sherry, S.~T., Zaranek, A.~W., Ball, M., Bobe, J., Estep, P., Church, G.~M.,
  Marks, P., Kyriazopoulou-Panagiotopoulou, S., Zheng, G.~X., Schnall-Levin,
  M., Ordonez, H.~S., Mudivarti, P.~A., Giorda, K., Sheng, Y., Rypdal, K.~B.,
  and Salit, M.
\newblock Extensive sequencing of seven human genomes to characterize benchmark
  reference materials.
\newblock \emph{Scientific Data}, 3:\penalty0 160025, Jun 2016.

\end{thebibliography}

    \pagebreak
    \begin{figure*}[!t]
        \centering
        \fbox{\parbox{0.35\textwidth}{\Large\bf\centering Supplementary Material}}
    \end{figure*}
    \appendix

\section{Extended Motivational Example}
Current schedulers prioritize the use of simple online heuristics~\cite{Grandl2016} and coarse-grained resource bucketing (e.g., core counts, free memory) and require user labeling of commonly used system resources~\cite{Hindman2011, Grandl2016:2} to make scheduling decisions.
Those approaches are untenable in truly heterogeneous settings as
\begin{enumerate*}[label=(\roman*)]
    \item defining such heuristics is difficult over the combinatorial space of application-processor/accelerator configurations; and
    \item user-based resource usage labeling requires in-depth understanding of the underlying system.
\end{enumerate*}
This paper demonstrates the use of ML to automatically infer such heuristics and their evolution over time as new user workloads and/or new accelerators are added.

\begin{figure}[!bp]
    \centering
    \includegraphics{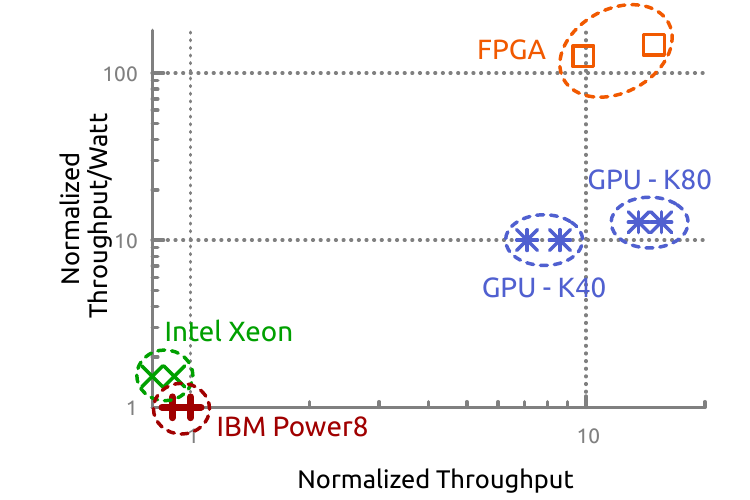}
    \caption{Architectural diversity leading to varied performance for the PairHMM kernel.}
    \label{fig:affinity}
\end{figure}

\subsection{Dealing with Architectural Heterogeneity}
We reiterate that state-of-the-art schedulers do not model the emergent heterogeneous compute platforms that are being widely adopted in data centers and hence leave a lot to be desired (as can also be seen in the performance of our baselines).
Consider, for example, the execution of the \emph{forward algorithm on PairHMM models}~\cite{Banerjee2017_FPL}, a computation that is commonly performed in computational genomics workloads.
\cref{fig:affinity} shows the significant diversity (nearly 100$\times$) in performance of this single workload across CPUs (from Intel and IBM), GPUs (two models of GPUs from NVIDIA) and FPGA implementations.
The increasing heterogeneity necessitates rethinking of the design and implementation of future schedulers, as the current approach will require an extraordinary amount of manual tuning and expertise to adapt to the emergent systems.
In contrast, the proposed technique eliminates that work and automates the whole process of learning the right granularity of resources and scheduling workloads in cloud-based, dynamic, multi-tenant environments, thereby improving application performance and system utilization, all with minimal human supervision.
Prior work uses microarchitectural throughput metrics such as clock cycles per instruction~\cite{Giceva2014, Delimitrou2013, Delimitrou2014, Mars2011, Mars2013} as proxies for processor affinities.
In our case, such metrics are not usable because of the wide diversity in processors, i.e., CPU-centric units cannot describe the performance of GPUs/FPGAs.

\begin{figure}[!t]
    \centering
    \includegraphics{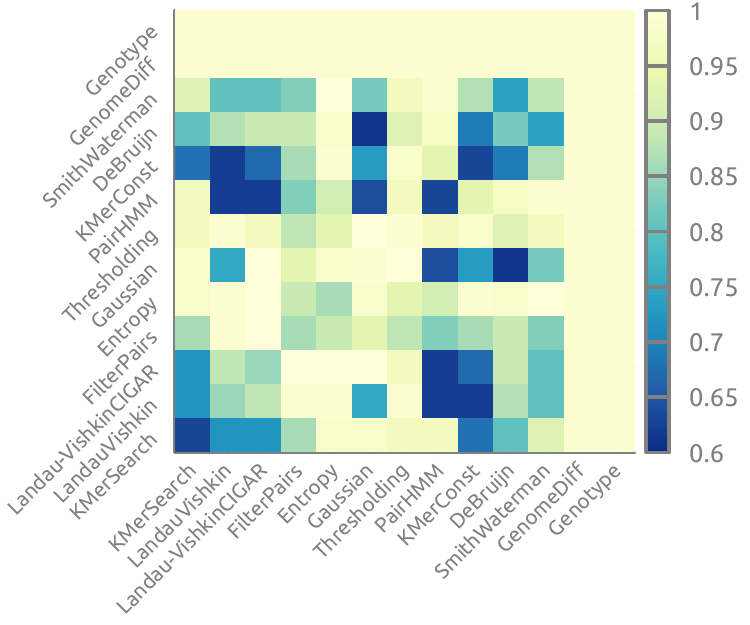}
    \caption{Degradation in runtime of co-located kernels due to shared resource contention.}
    \label{fig:interference}
\end{figure}

\subsection{Dealing with Resource Granularity}
Traditional schedulers use coarse-grained resource bucketing, i.e., they schedule macro-resources like CPU core counts and GBs of memory.
That simplifies the design of the scheduling algorithms (both the optimization algorithms and attached heuristics), resulting in an inability to measure low-level sources of resource contention in the system.
The contention of such low-level resources is often the cause for performance degradation and variability.
Consider, for example, the concurrent execution of several compute kernels (described in Appendix~\ref{sec:workloads}) on co-located hyper-threads (i.e., threads that share resources on a single core) on an Intel CPU.
If we abstract the problem at the level of CPU threads and memory allocated, then those kernels should execute in isolation.
The normalized runtime variation is illustrated in \cref{fig:interference}.
We observe a slowdown of as much as 40\% (i.e., the co-located runtime is 60\% of the isolated runtime) for some combinations of kernels, and almost no slowdown for others.
That problem is further exacerbated by the architectural diversity in processors that we described earlier.
The proposed technique accounts for such contention by explicitly collecting information on low-level system state by using performance counter measurements, and by estimating resource usage in the system by explicitly encoding the measurements in its POMDP model.

\begin{figure*}[!t]
    \centering
    \includegraphics{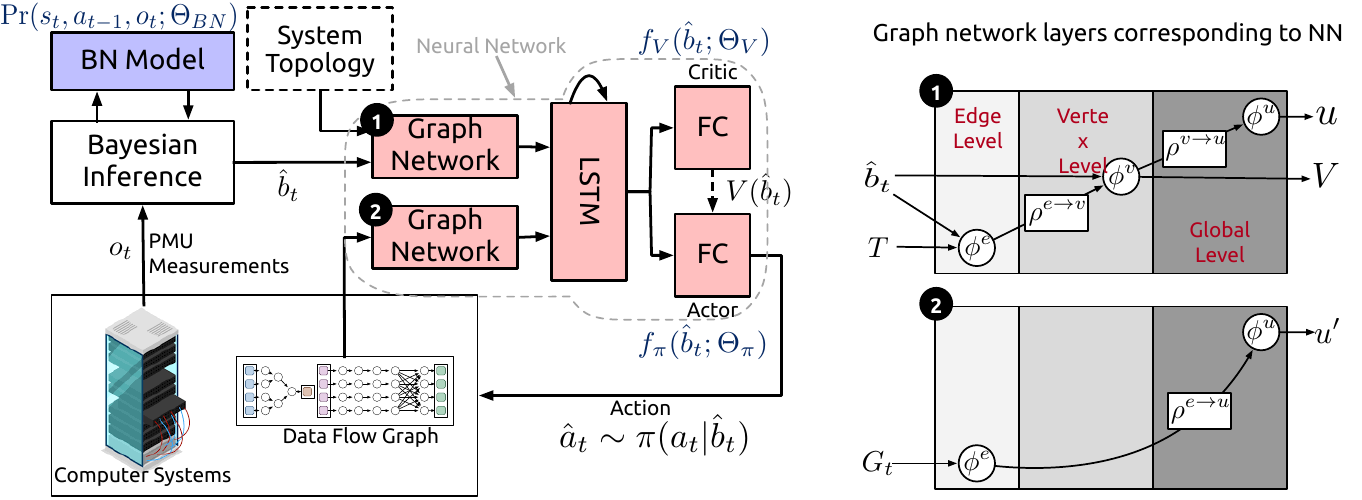}
    \caption{Proposed POMDP model.}
    \label{fig:dnn}
\end{figure*}

\section{Implementation Details}
The scheduling framework functions as follows.
\begin{enumerate}[noitemsep,nolistsep,leftmargin=*]
    \item The scheduler first makes measurements by using the available processor performance counters (e.g., instructions retired, cache misses).
    \item When a processor becomes idle (finishes running the current kernel), it invokes the scheduler.
    \item The measurements are fed into the scheduler's BN model as input.
    Using those measurements, the BN model computes the utilization of different levels of architectural resources in the system (e.g., memory bandwidth utilization, PCIe link utilization).
    We refer to those utilizations as the \textit{state} of the system.
    \item The computed utilization numbers, user programs represented as a DFG, and a system topology graph are fed into an NN. The NN produces a scheduling decision that is actuated in the system. The action space consists of a kernel-processor pair.
    \item Finally, the scheduler gets feedback from the system (i.e., the reward) in terms of the time it took for the job to run as a result of its scheduling decision.
    \item While in \textit{training mode}, if an incorrect decision is made, Symphony enqueues an update of the policy parameters using back-propagation on the A2C/A3C loss function.
    An incorrect decision is one where kernel input-output dependencies are not respected, or a kernel-accelerator pair is picked where the accelerator does not provide an implementation of the kernel.
\end{enumerate}

\begin{table}[!t]
    \centering
    \caption{Mapping of the graph network layer functions in \cref{fig:dnn}. We use the notation $FCNN(a, b)$ to denote a 2-hidden fully-connected layers with $a$ and $b$ hidden units, respectively.}
    \label{tab:gn}
    \begin{tabular}{rll}
        \toprule
        Function in GN & Function in \circled{1} & Function in \circled{2} \\
        \midrule
        $\phi^e$ & $FCNN(64,32)$ & $FCNN(64,32)$\\
        $\phi^v$ & $FCNN(32,16)$ & --\\
        $\phi^u$ & $FCNN(16,16)$ & $FCNN(32,16)$\\
        $\rho^{e\rightarrow v}$ & $\sum e$ & --\\
        $\rho^{v\rightarrow u}$ & $\sum v$ & --\\
        $\rho^{e\rightarrow u}$ & -- & $ReLU(e)$\\
        \bottomrule
    \end{tabular}%
\end{table}

\subsection{Graph Network Details}
The structure of the graph network used in the proposed model is illustrated in \cref{fig:dnn}.
The numbers of parameters used in the different layers of the graph network are listed in \cref{tab:gn}.

\subsection{Hyperparameters}
The hyperparameters used to train the proposed POMDP model are listed in \cref{tab:hyper}.

\begin{table}[!bp]
    \centering
    \caption{Hyperparameters used in the model.}
    \label{tab:hyper}
    \begin{tabular}{ll}
        \toprule
        \textbf{Hyperparameter} & \textbf{Value} \\
        \midrule
        Learning Rate & 0.005 \\
        LSTM Unroll Length & 20 \\
        $n_s$ & 20 \\
        $n_e$ & 2 \\
        \bottomrule
    \end{tabular}%
\end{table}

\subsection{System Measurement Details}
\begin{figure}[!t]
    \centering
    \includegraphics[width=\columnwidth]{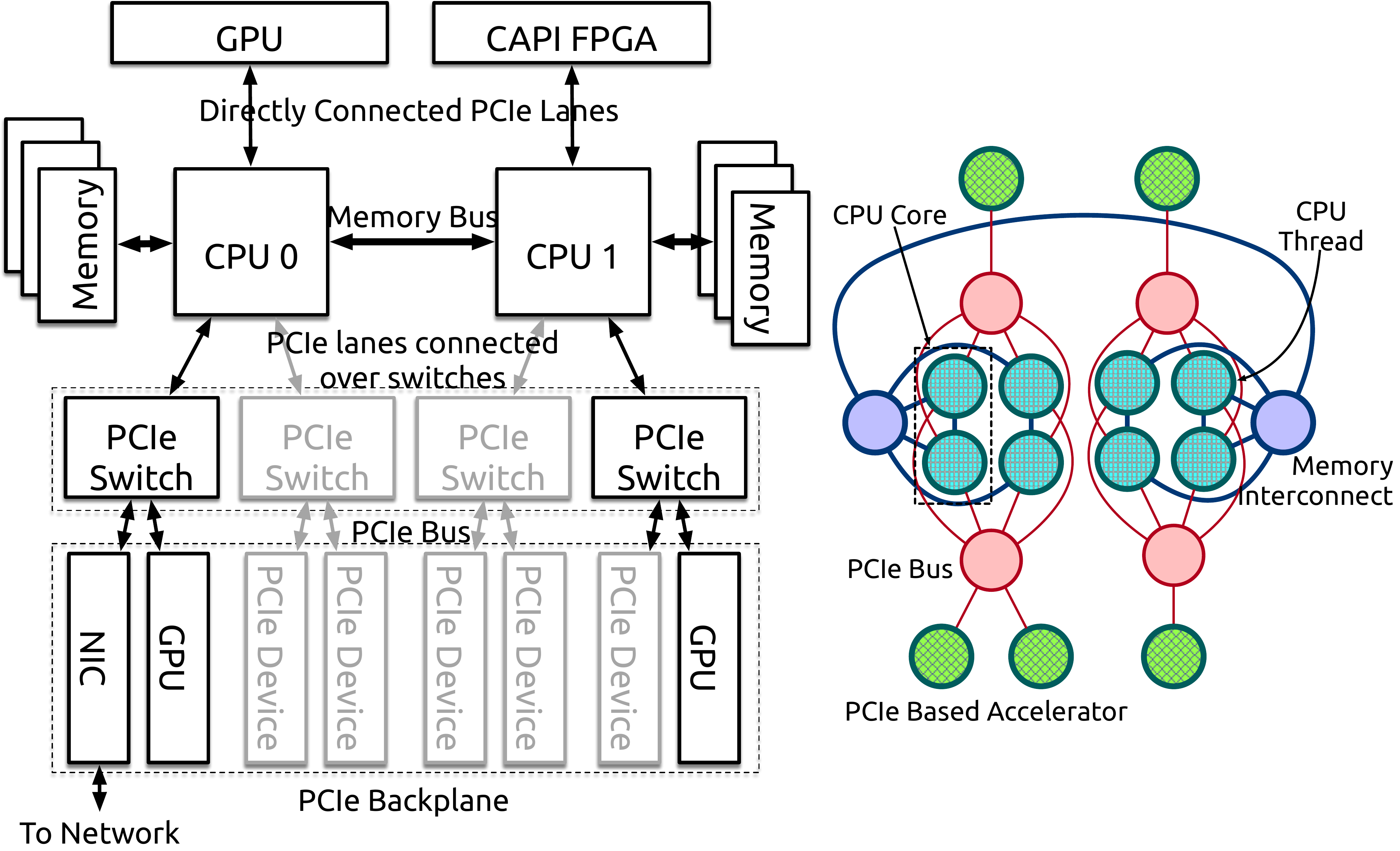}
    \caption{Example of a dual-socket NUMA-based system topology with a PCIe-interconnect and -devices. Figure on the right shows an graph-encoding of the topology.}
    \label{fig:topology}
\end{figure}

\textbf{Topology Information.}
Consider the example of standard NUMA based computing system with PCIe based accelerators shown in \cref{fig:topology}.
The system contains
\begin{enumerate*}[label=(\roman*)]
    \item multiple CPUs which have non-uniform access to memory,
    \item several accelerators (including GPUs and FPGAs) each with their own memory, and
    \item a system interconnect which connects all of the components of the system together.
\end{enumerate*}
Symphony encodes the system topology as a graph $T = (P, N)$ (also shown in \cref{fig:topology}).
The nodes of the graph $P$ correspond to the processing elements (and attached memory) and memory/system interconnects.
Each of the these nodes $p \in P$ have an attached resource utilization vector. For example, in an Intel processor, the utilization vector would include utilization like that of micro-op issue ports, floating point unit utilization etc.~\cite{IntelHotChips, IntelAORM}.

The scheduler queries the system topology and builds the topology graph $T$ (which is used as an input to the RL agent) using hwloc~\cite{Broquedis2010}.
hwloc provides information about CPU cores, caches, NUMA memory nodes, and the PCIe interconnect layout (i.e., connections between the PCIe root complex and PCIe switches), as well as connection information on peripheral accelerators, storage, and network devices in the system.
The scheduler does not explicitly model the rack-scale or data center network (unlike some previous approaches, e.g.,~\citet{Isard2009, Chowdhury2014}), but the BN and RL model can be extended to do so.
Our measurements considers injection bandwidth at the network interface card (NIC) to be a proxy for network performance, i.e., the NIC is modeled as an accelerator that accepts data at $\min(\text{PCIe Bandwidth, Injection Bandwidth})$.

\textbf{Performance Counter Measurements.}
Performance counters' configuration and access instructions require kernel mode privileges, and hence those operations are supported by Linux: system calls to configure and read the performance counter data.
Symphony uses a combination of user-space tools, e.g., libPAPI~\cite{Terpstra2010}, PMUTools~\cite{PMUTools}, and perf that wrap around the system call interface to make both system-specific and system-independent measurements.

We configure the performance counters to make system-wide measurements (i.e., for all processes).
If the performance counter measurements are configured in that way, it might incur security risks, particularly by opening up side channels through which attackers could infer workload characteristics.
However, analysis or mitigation of such risks is not in the scope of this paper and may form the basis of future work.

All kernel executions are non-preemptive in the context of the proposed runtime, however the OS scheduler can preempt CPU threads.
Further we prevent the OS scheduler from re-balance tasks/threads once assigned to a particular CPU.
This is achieved by explicitly setting affinities of threads to cores (i.e., pinning them).

\textbf{Performance Penalties.}
Monitoring of performance counters without having to perform interrupts is almost free.
In our implementation, we capture on-core performance counters directly before and after a single kernel invocation.
Un-core performance counters are measured periodically (every million dynamic instructions on a core) by using a \emph{performance monitoring interrupt}.
On an IBM PowerPC processor, the interrupt handler initiates a DMA transfer of the performance counters to memory~\cite{IBMDMAPerf}, thereby incurring no performance penalty (other than the time to service the interrupt). On Intel processors, the interrupt handler has to explicitly read the performance counter registers and write them to memory.
In our tests (on Intel processors), we observed a \textasciitilde 3\% performance penalty for applications with interrupts enabled.
That corresponds to an execution of a usermode interrupt with an average 900-ns latency.

\textbf{Distributed Execution.}
In our evaluation we have deployed Symphony in a rack-scale distributed context (over an EDR Infiniband network fabric) as a centralized scheduler controlling all processing resources.
Here, all the performance counter measurements are sent over the network to a centralized server that makes scheduling decisions.
This approach works well at the scale of a rack, where all resources are essentially one hop away at 0.2-\textmu{}s latency.
Extending Symphony to larger or slower networks might present challenges, where network latency causes stale performance counter data to reach the scheduler.
We will address these challenges in future work.

\subsection{Dynamically Reconfigurable FPGA Accelerator}

\begin{figure}[!t]
    \centering
    \includegraphics[width=\columnwidth]{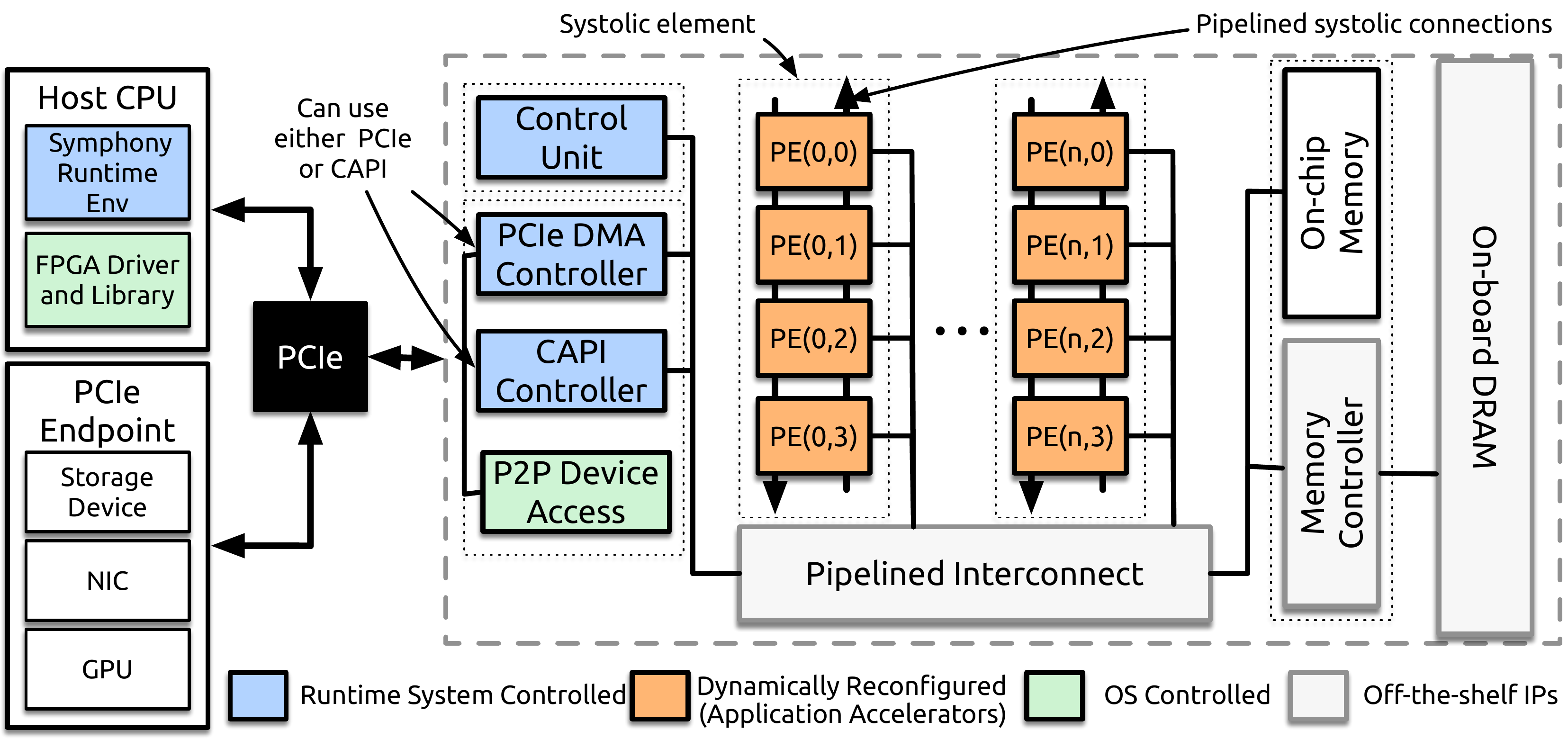}
    \caption{Architecture of the FPGA-based hardware co-processor controlled by Symphony.}
    \label{fig:tcga_architecture}
\end{figure}

Our implementation and evaluation of Symphony uses a custom FPGA accelerator (see \cref{fig:tcga_architecture}).
Due to space limitations, here we briefly describe the features of the accelerator.
\begin{itemize}[noitemsep,nolistsep,leftmargin=*,nosep]
    \item \emph{Processing Elements (PEs).}
    The co-processor is optimized to execute the computational kernels as a single instruction of the application.
    Sets of four neighboring PEs are directly connected as a systolic element, thereby enabling high bandwidth data transfer in between PEs and forming the quantum of reconfiguration.

\begin{table}[!t]
    \centering
    \caption{Hardware specifications of test cluster.}
    \begin{tabular}{crp{5cm}}
        \toprule
        \textbf{Name} & \textbf{\#} & \textbf{Specifications}\\
        \midrule
        \texttt{M1} & 2 & CPU IBM Power8 (SMT 8); 870 GB RAM; GPU NVIDIA K80; FPGA Alpha
        Data 7V3\\
        \texttt{M2} & 4 & CPU IBM Power8 (SMT 4); 512 GB RAM; GPU NVIDIA K40; FPGA
        Nallatech 385\\
        \texttt{N} & 1 & Mellanox FDR Infiniband\\
        \bottomrule
    \end{tabular}
    \label{tab:cluster}
\end{table}

\begin{table*}[!bp]
    \centering
    \caption{Enumeration of workloads used to evaluation.}
    \begin{tabular}{lcccp{9cm}}
        \toprule
        \textbf{Application} & \multicolumn{3}{c}{\textbf{Processors}} & \textbf{Implementations}\\ \cmidrule{2-4}
        & \textbf{CPU} & \textbf{GPU} & \textbf{FPGA} &\\  \midrule
        Alignment \emph{(Align)} & \cmark & \cmark & \cmark & \cite{Li2009BWA, Li2010, Langmead2009, Zaharia2011snap, Banerjee2018_ASAP, Banerjee2016}, \\
        Indel Realignment \emph{(IR)} & \cmark & \xmark & \xmark & \cite{McKenna2010, Nothaft2015} \\
        Variant Calling \emph{(HC)} & \cmark & \cmark & \cmark & \cite{Li2009, McKenna2010, Nothaft2015ms, Rimmer2014, Banerjee2017_FPL} \\
        EEG-Graph \emph{(EEG)} & \cmark & \cmark & \cmark &  \cite{Varatharajah2017, Banerjee2019}\\
        AttackTagger \emph{(AT)} & \cmark & \cmark & \cmark &  \cite{Cao2015, Banerjee2019}\\
        \bottomrule
    \end{tabular}
	\label{tab:kernels}
\end{table*}

    \item \emph{Host-FPGA Communication.}
    The board interfaces with the host CPU over PCIe and can be configured to communicate with the host processor over this interface in one of two ways:
    \begin{enumerate*}[label=(\roman*)]
        \item using direct memory access (DMA) to the hosts memory over the PCIe bus, or
        \item using IBM's coherent accelerator processor interface (CAPI)~\cite{CAPI2015}.
    \end{enumerate*}
    \item \emph{Dynamic Reconfiguration.}
    The configuration of the accelerator (i.e., which kernels PEs are available at any time) is controlled by Symphony.
    Symphony treats the reconfiguration of the accelerator as a kernel that has to be dispatched to the FPGA.
    The state of the accelerator is fed into Symphony along with the system topology $T$.
    \item \emph{Launching Kernels.}
    Remember CPU executors (i.e., threads which are given tasks to execute) are pinned or bound to underlying hardware SMT thread.
    Accelerators however require the CPUs to initiate their execution. As a result, each accelerator in the system is assigned a proxy executor thread that orchestrates (i.e., launches, polls for completion etc.) its execution.
    These executors are responsible for managing their own queues for maintaining tasks that are ``waiting'' for execution.
\end{itemize}

\section{Evaluation Environment}

\subsection{Evaluation System}
All evaluation experiments are performed on an 11 node rack-scale test-bed of IBM Power8 CPUs, NVIDIA K40 and K80 GPUs, as well as FPGAs (listed in \cref{tab:cluster}).
All the machines in the cluster are connected using a single switch EDR Infiniband network.

\subsection{Evaluation Workloads}
\label{sec:workloads}

We illustrated the generality of the proposed approach on a variety of real-world workloads (listed in \cref{tab:kernels}) that used CPUs, GPUs, and FPGAs:
\begin{enumerate}[nosep]
    \item \emph{variant-calling and genotyping analysis}~\cite{VanDerAuwera2013} on human genome datasets appropriate for clinical use (consisting of \texttt{Align}, \texttt{IR}, and \texttt{HC} in \cref{tab:kernels}),
	\item \emph{epilepsy detection and localization}~\cite{Varatharajah2017} on intra-cranial electroencephalography data; and
	\item online \emph{security analytics}~\cite{Cao2015} on network- and host-level intrusion detection system event-streams.
\end{enumerate}
For the variant-calling and genotyping workload we use the NA12878 genome sample from the GIAB consortium~\cite{Zook2016} for all our experiments as it is representative of human clinical datasets.
For the EEG and AT workloads, we use the same datasets as discussed in the original papers.

\end{document}